\documentclass[11pt,a4paper]{article}
\usepackage[utf8]{inputenc}

\usepackage{jheppub}
\usepackage{braket}
\usepackage{comment}
\usepackage{amsmath,amsfonts,amssymb,bbm, mathtools, mathrsfs}

\usepackage{csquotes}
\usepackage{graphicx, color}
\usepackage{tikz,pgf,tikz-cd,float}
\usepackage{physics,tensor}
\usepackage{comment}
\usepackage{ulem}

\newcommand{\diff}{\mathrm{d}}
\newcommand{\ee}{\mathrm{e}}
\newcommand{\ii}{\mathrm{i}}
\newcommand{\qmarks}[1]{``#1''}
\newcommand{\one}{\mathbbm 1}

\newcommand{\bg}[1]{\bar{#1}_{j}} 
\newcommand{\bgs}[1]{\bar{#1}_{j_{o}}} 
\newcommand{\fo}[1]{\delta #1 _{j}} 
\newcommand{\foS}[1]{\delta #1 _{\text{S},j}} 
\newcommand{\foT}[1]{\delta #1 _{\text{T},j}} 

\newcommand{\fos}[1]{\delta #1 _{j_{o}}} 

\newcommand{\Conv}{\mathop{\scalebox{2}{\raisebox{-0.2ex}{$\ast$}}}}%

\newcommand{\chiz}{\chi_{0}} 
\newcommand{\chimz}{\pi_{0}} 
\newcommand{\chimzt}{\tilde{\pi}_{0}} 
\newcommand{\psim}{\pi_{\psi}} 
\newcommand{\psimz}{\pi_{\psi 0}} 

\title{Interacting Scalar Field Cosmology from Full Quantum Gravity}
\author[a]{Tom Ladst\"atter}
\emailAdd{tomr.ladstaetter@proton.me}
\author[b]{Luca Marchetti}
\emailAdd{luca.marchetti@oist.jp}
\author[]{}

\affiliation[a]{Arnold Sommerfeld Center for Theoretical Physics, Ludwig-Maximilians-Universit{\"a}t M{\"u}nchen,\\ Theresienstrasse 37, 80333 M{\"u}nchen, Germany}
\affiliation[b]{Okinawa Institute of Science and Technology Graduate University,\\
Onna, Okinawa 904 0495 Japan}
\affiliation[b]{
Kavli Institute for the Physics and Mathematics of the Universe (WPI),\\ UTIAS, The University of Tokyo, Chiba 277-8583, Japan
}
\affiliation[]{}

\emailAdd{}
\emailAdd{}

\abstract{
We study the relational cosmological dynamics emerging from interacting group field theory (GFT) models minimally coupled to a massless clock scalar field and a self-interacting scalar field. We focus on two broad classes of GFT interactions—pseudosimplicial and pseudotensorial—which generalize simplicial and tensorial interactions, respectively. Treating these interactions perturbatively, we extract the effective cosmological dynamics using mean-field techniques.
In the geometric sector, we identify appropriate classical limits of the resulting dynamics, characterized by the emergence of a cosmological constant term in pseudotensorial models and of dynamical dark energy in pseudosimplicial ones. In the matter sector, we find that quantum gravity interactions induce a mass term and modify the classical symmetries of the scalar field dynamics, allowing for a consistent classical matter–geometry description only for specific forms of the effective scalar field potential. Finally, we show that these quantum gravity compatibility conditions on the effective potentials can be relaxed by allowing for a scale-dependent gravitational coupling, and that this running is uniquely fixed once the classical scalar field potential is specified.
}


\arxivnumber{}

\begin{document}
\maketitle
\flushbottom

\section{Introduction}\label{sec:introduction}
The $\Lambda$CDM model provides an exceptionally accurate description of our universe, successfully accounting for a wide range of cosmological observations \cite{Planck:2018vyg}. However, despite its many successes, recent tensions—such as the discrepancy in the measured values of the Hubble constant $H_0$ and a distinct preference for a dynamical dark energy—have emerged \cite{DiValentino:2021izs,DESI:2025zgx}, suggesting that $\Lambda$CDM may not be sufficiently rich to fully capture the fundamental properties of our universe. This is not unexpected, as $\Lambda$CDM is an effective model that lacks ultraviolet (UV) completion. Such a completion would be essential to gaining a deeper understanding of key open questions, including the nature of dark energy, dark matter, and (to some extent) the origin of primordial perturbations. Quantum gravity  provides a promising avenue to address these gaps by providing a UV-complete framework for cosmology, potentially offering insights into these foundational questions. At the same time, cosmology itself may serve as a unique testing ground for quantum gravity, as traces of quantum gravitational effects could be imprinted in the early universe.

However, extracting cosmological physics from quantum gravity is a highly non-trivial task. In background-independent approaches to quantum gravity, non-perturbative quantization often leads to theories in which conventional spacetime structures are significantly weakened or even entirely absent, in which case the fundamental degrees of freedom may take a purely pre-geometric form. In such cases, the key challenges are to identify~\cite{Oriti:2018dsg}, in appropriate limits (i) \qmarks{macroscopic} structures that can be associated with continuous and (semi-)classical observables, and (ii) effective dynamical equations for these quantities that, in these limits, recover the general relativistic ones. Note that both these emergent macroscopic quantities and their dynamics naturally encode gauge (i.e. diffeormophism)-invariant information, particularly when derived from an inherently background-independent theory. This naturally suggests a relational perspective, where physical descriptions are constructed with respect to a dynamical— and possibly quantum—reference frame~\cite{Rovelli:1990ph,Hoehn:2019fsy,Rovelli:2001bz,Dittrich:2005kc,Goeller:2022rsx}.

These challenges find a concrete realization—and, to some extent, concrete resolutions—within the tensorial group field theory (TGFT)~\cite{Freidel:2005qe,Oriti:2006se,Carrozza:2013oiy,Carrozza:2016vsq} approach to quantum gravity and its cosmological applications \cite{Gielen:2013kla,Oriti:2016acw,Gielen:2016dss,Marchetti:2020umh,Jercher:2021bie}.
TGFTs are field-theoretic generalizations of matrix and tensor models~\cite{DiFrancesco:1993cyw,Gurau:2011xp,GurauBook,Gurau:2016cjo,Jercher:2022mky}
. When endowed with quantum geometric degrees of freedom encoded in group-theoretic data, these theories are known as Group Field Theories (GFTs), and can be understood as quantum and statistical field theories of spacetime defined on a group manifold. Their quantum geometric nature establishes deep links to other approaches to quantum gravity, such as loop quantum gravity (LQG)~\cite{Ashtekar:2004eh,Oriti:2013aqa,Oriti:2014yla}, spin foam models~\cite{Perez:2003vx,Perez:2012wv}, simplicial gravity~\cite{Bonzom:2009hw,Baratin:2010wi,Baratin:2011tx,Baratin:2011hp,Finocchiaro:2018hks} or dynamical triangulations~\cite{Loll:1998aj,Ambjorn:2012jv,Jordan:2013sok,Loll:2019rdj}. 

In cosmological applications of GFTs, the aforementioned macroscopic quantities and effective dynamical equations are associated with averaged collective operators and quantum field equations on coherent peaked states (CPS). These states not only capture mean-field~\cite{Marchetti:2020xvf, Marchetti:2022igl,Marchetti:2022nrf} collective behavior but also allow for a relational interpretation of the resulting averages \cite{Marchetti:2020umh,Marchetti:2020qsq}. The emergent cosmological dynamics successfully reproduces the Friedmann equations for spatially homogeneous and isotropic flat geometries at late relational times, while at early relational times the initial Big Bang singularity of classical cosmology is replaced by a quantum bounce \cite{Oriti:2016qtz,Gielen:2019kae,Marchetti:2020umh,Jercher:2021bie}.
At the inhomogeneous level, recent results—improving upon previous attempts \cite{Gielen:2017eco,Gerhardt:2018byq,Marchetti:2021gcv}—have shown that cosmological perturbations naturally arise from entanglement among the underlying quantum geometric degrees of freedom. Moreover, their dynamics match the general relativistic ones at late times for sub-Planckian modes while exhibiting corrections at trans-Planckian scales \cite{Jercher:2023kfr,Jercher:2023nxa}.

It is important to note, however, that these results are derived under the assumption of a matter content consisting solely of minimally coupled, massless, and free scalar fields, and in the limit where quantum gravitational interactions are negligible. For cosmological applications, it is essential to consider a more realistic matter content. In this regard, a natural step is to investigate the impact of a scalar field potential, as self-interacting scalars may have played a crucial role in the early universe.
Moreover, quantum gravitational interactions are expected to become significant at late times, eventually dominating the underlying quantum gravity dynamics. While previous studies have explored the cosmological consequences of GFT interactions \cite{deCesare:2016rsf,Oriti:2025lwx}, revealing an intriguing connection between such interactions and emergent acceleration, they have primarily focused on simplified models that capture only certain aspects of more realistic GFT interactions.

In this work, we address both of these issues by investigating the cosmological properties of a GFT model describing gravity coupled to a self-interacting scalar field \cite{Li:2017uao}. The physical content of such a GFT model is encoded in the structure of the perturbative Feynman amplitudes, which take the form of lattice gravity path-integrals including discretized scalar field degrees of freedom (see \cite{Ambjorn1992_3d,HamberWilliams1994,Ambjorn1993_4d,MoralesTecotlRovelli1995,Thiemann1998_QSD5,OritiPfeiffer2002,FreidelLouapre2004,FreidelOritiRyan2005,OritiRyan2006,Fairbairn2007,Speziale2007,DowdallFairbairn2011,Dowdall2011_WilsonLoops,BianchiHanRovelliWielandMagliaroPerini2013,LewandowskiSahlmann2016, Jercher:2024hlr} for earlier works on matter coupling in discrete path-integrals, LQG, spinfoams and GFTs). This directly informs the construction of the GFT action, and in particular, the form of its interaction kernel, where  features of the scalar field potential are naturally encoded. As a result, our study requires a systematic analysis of the role of GFT interactions in shaping the emergent cosmological dynamics. Crucially, while the classical dynamical properties of the gravitational and scalar fields are determined by the \textit{perturbative} structure of the GFT model, the resulting \textit{emergent} continuum dynamics, which capture inherently \textit{non-perturbative} effects, may differ significantly from their classical counterparts. In particular, they may depend on the various choices and approximations—such as the selection of the quantum state and the use of mean-field methods—that are necessary to define and study the emergent limit.

As we shall see, the interacting GFT mean-field equations describing such emergent behavior form a system of coupled, highly nonlinear differential equations. These are extremely difficult to solve analytically. Therefore, in this work, we restrict our analysis to the regime of small interactions and seek correspondingly perturbative solutions. This approach allows us to study a very broad class of GFT interactions, which we classify as pseudotensorial and pseudosimplicial interactions, following the terminology introduced in \cite{deCesare:2016rsf}.

In the pseudotensorial case, we show that matching with classical Friedmann dynamics is possible for interactions of a specific order, and that, in this case, a cosmological constant naturally emerges. This leads to a corresponding emergent mass term in the effective scalar field dynamics, which is positive for a negative cosmological constant and vice versa. Furthermore, we demonstrate that if the gravitational coupling $G$ is not running, only a specific class of (technically natural) scalar field potentials is compatible with the quantum gravity-induced effective scalar field dynamics. Finally, we discuss how these quantum gravity-compatibility conditions can be relaxed by allowing a running gravitational coupling.

Similar results hold for the pseudosimplicial case, where classical dynamics—both in the gravitational and matter sectors—are matched only for interactions of a specific order (identical to the pseudotensorial case) and for a specific class of scalar field potentials (distinct from the pseudotensorial case) if the gravitational coupling is not running. Importantly, we show that in this case, quantum gravity effects break any shift symmetry potentially present in the original classical action down to a discrete subgroup, which is reminiscent of non-perturbative effects in axion physics \cite{GrillidiCortona:2015jxo}. Analogously, the emergent dark energy component takes a non-trivial time dependence in this case. Finally, we discuss how these quantum gravity-compatibility conditions can be relaxed by allowing a running gravitational coupling.

The structure of this paper is as follows. In Sec.\ \ref{sec:review}, we introduce the GFT approach to quantum gravity, focusing on its action and the structure of its interaction terms, particularly in the pseudosimplicial and pseudotensorial cases. In Sec.\ \ref{sec:variation}, we derive the resulting mean-field equations and explicitly solve them perturbatively. In Sec.\ \ref{sec:averages}, we use these equations to obtain the corresponding equations for geometric (volume) and matter (scalar field) relational observables. In Sec.\ \ref{sec:emergentscalarfieldcosmology}, we compare these equations with their classical counterparts and discuss matching conditions as well as intrinsic differences. Finally, in Sec.\ \ref{sec:conclusion}, we present our conclusions and outline directions for future work.
\section{The GFT approach to quantum gravity}\label{sec:review}
In the GFT approach to quantum gravity, one constructs a generating functional for the discrete gravitational path-integral. This generating functional takes the form of a partition function for a GFT field $\Phi:\mathcal{D}\to \mathbb{C}$ with action $S_{\text{GFT}}$. Indeed, by choosing the group domain $\mathcal{D}$ and $S_{\text{GFT}}$ appropriately, one can show \cite{Freidel:2005qe}: (i) that the corresponding Feynman graphs $\Gamma$ are dual to cellular complexes with arbitrary topology, and (ii) that the quantum amplitudes $\mathcal{A}_\Gamma$ associated with $\Gamma$ can be identified with (first-order) simplicial gravity path-integrals on a discrete structure dual to the graph $\Gamma$. If, instead of representing discrete gravitational quantities in terms of group theoretic data one uses representation data, the same amplitudes $\mathcal{A}_\Gamma$
 can be identified with appropriate spinfoam models. Note that as the GFT field is defined on a field space, rather than spacetime, spacetime notions can only be emergent within this framework. 

In the following we will focus on a GFT associated with the EPRL model \cite{Engle:2007wy, Freidel:2007py} minimally coupled with $m$ massless scalar fields \cite{Oriti:2016qtz, Li:2017uao, Marchetti:2021gcv}. In this case, $\mathcal{D}=\mathrm{SU}(2)^4\times\mathbb{R}^m$; furthermore, one imposes the following closure constraint \cite{Oriti:2011jm}:
\begin{equation}\label{eqn:closureconstraint}
    \Phi(\vec{g},\boldsymbol{\chi})=\Phi(\vec{g}\cdot_{\text{d}}h ,\boldsymbol{\chi})\,,\qquad \forall \, h\in\mathrm{SU}(2)\,,
\end{equation}
where $\vec{g}\cdot_\text{d} h\equiv \{g_1h,\dots,g_4h\}$ represents the diagonal right action of $h$ on $\vec{g}\equiv (g_1,\dots,g_4)\in\mathrm{SU}(2)^4$, and $\boldsymbol{\chi}=\{\chi^1,\dots,\chi^m\}\in\mathbb{R}^m$. Due to equation \eqref{eqn:closureconstraint}, the quanta of the theory can be associated with $3$-simplices (tetrahedra), or, equivalently, as open spin-networks \cite{Oriti:2013aqa, Oriti:2014yla} whose links are decorated with the equivalence class of geometric data (holonomies) $\{\vec{g}\}\equiv \{\vec{g}\cdot_{\text{d}}h,h\in\mathrm{SU}(2)\}$, and whose nodes are decorated with scalar field values $\boldsymbol{\chi}$. This interpretation is even more manifest in the spin representation
\begin{align}\label{eqn:grouptospin}
    \Phi(\vec{g},\boldsymbol{\chi})&=\sum_\iota\sum_{\vec{j}}\sum_{\vec{m},\vec{n}}\Phi^{j_1,\dots,j_4;\iota}_{m_1,\dots,m_4}(\boldsymbol{\chi})\left[\prod_{i=1}^4\sqrt{d(j_i)}D^{j_i}_{m_in_i}(g_i)\right]\mathcal{I}^{j_1,\dots,j_4;\iota}_{n_1,\dots,n_4}\nonumber\\
    &\equiv \sum_{\vec{\kappa}}\Phi_{\vec{\kappa}}(\boldsymbol{\chi}) \, \psi_{\vec{\kappa}}(\vec{g})\,,
\end{align}
where $D^{j}_{mn}(g)$ are Wigner functions, $d(j)=2j+1$, and $\mathcal{I}^{j_1,\dots,j_4;\iota}_{n_1,\dots,n_4}$ is an $\mathrm{SU}(2)$ intertwiner that appears because the right-invariance imposed by \eqref{eqn:closureconstraint} requires $\Phi$ to be expanded on a basis of functions of $L^2(G^4/G)$. Field modes are therefore labeled by spin network data $\vec{\kappa}\equiv \{\vec{j},\vec{m},\iota\}$, where $\vec{j}$ and $\vec{m}$ are respectively spin and angular momentum labels, while $\iota$ is an intertwiner quantum number.
\subsection{The GFT Fock space}
\label{sec:gftfock}
GFTs can be formulated in a Fock language \cite{Gielen:2024sxs} by introducing 
 creation $\hat{\varphi}^\dagger(\vec{g},\boldsymbol{\chi})$ and annihilation $\hat{\varphi}(\vec{g},\boldsymbol{\chi})$ operators that satisfy the following commutation relations:
\begin{subequations}\label{eqn:commutationrelationsg}
    \begin{align}
        [\hat{\varphi}(\vec{g},\boldsymbol{\chi}),\hat{\varphi}^\dagger(\vec{g}^{{}\,\prime},\boldsymbol{\chi}')]&=\one_G(\vec{g},\vec{g}^{{}\,\prime})\delta(\boldsymbol{\chi}-\boldsymbol{\chi}')\,,\\
        [\hat{\varphi}(\vec{g},\boldsymbol{\chi}),\hat{\varphi}(\vec{g}^{{}\,\prime},\boldsymbol{\chi}')]&=[\hat{\varphi}^\dagger(\vec{g},\boldsymbol{\chi}),\hat{\varphi}^\dagger(\vec{g}^{{}\,\prime},\boldsymbol{\chi}')]=0\,,
    \end{align}
\end{subequations}
where $\one_G$ is the Dirac delta distribution in the space $G^4/G$, with $G=\mathrm{SU}(2)$. Equivalently, in terms of representation labels, one can write
\begin{subequations}\label{eqn:commutationrelationsk}
    \begin{align}
        [\hat{\varphi}_{\vec{\kappa}}(\boldsymbol{\chi}),\hat{\varphi}^\dagger_{\vec{\kappa}'}(\boldsymbol{\chi}')]&=\delta_{\vec{\kappa},\vec{\kappa}'}\delta(\boldsymbol{\chi}-\boldsymbol{\chi}')\,,\\
        [\hat{\varphi}_{\vec{\kappa}}(\boldsymbol{\chi}),\hat{\varphi}_{\vec{\kappa}'}(\boldsymbol{\chi}')]&=[\hat{\varphi}^\dagger_{\vec{\kappa}}(\boldsymbol{\chi}),\hat{\varphi}_{\vec{\kappa}'}^\dagger(\boldsymbol{\chi}')]=0\,.
    \end{align}
\end{subequations}
In the following, we will identify $\hat\Phi_{\vec{\kappa}}(\boldsymbol{\chi}) = \hat \varphi_{\vec{\kappa}}(\boldsymbol{\chi})$ and $\hat \Phi^\dagger_{\vec{\kappa}}(\boldsymbol{\chi}) = \hat\varphi^\dagger_{\vec{\kappa}}(\boldsymbol{\chi})$ \cite{Oriti:2013aqa}, although other options are possible, especially if the field is real \cite{Marchetti:2024tjq}. As in standard QFT, a Fock space $\mathcal{F}$ can be constructed by the repeated action of the creation operator $\hat{\varphi}^\dagger_{\vec{\kappa}}(\boldsymbol{\chi})$ on the Fock vacuum $\ket{0}$. This Fock representation can be shown to be equivalent to a path-integral one (in turn strictly related to simplicial gravity and spinfoam models, as explained above) for a GFT quantization of the Husain-Kucka\v{r} model \cite{Marchetti:2024tjq}. This, together with the results of \cite{kegeles}, seems to suggest that (in perfect analogy with standard QFTs) the Fock structure for interacting GFTs would differ from the one given above  (see also \cite{Kegeles:2017ems} for non-Fock representations of interacting GFTs). However, since in the following we will introduce interactions only perturbatively, we will work within the above Fock space $\mathcal{F}$, (once again in perfect analogy with standard QFT). 
\paragraph{States.}
A one-particle state in $\mathcal{F}$ takes the following form:
\begin{equation}
    \ket{\vec{\kappa};\boldsymbol{\chi}}\equiv\hat{\varphi}^\dagger_{\vec{\kappa}}(\boldsymbol{\chi})\ket{0},
\end{equation}
and, as mentioned above, is therefore naturally associated with an open spin-network vertex with spin labels $\vec\kappa = (\iota, j_i, m_i)$ and scalar field data $\boldsymbol{\chi}=(\chi^1,\dots,\chi^m)$. Note that generic states in $\mathcal{F}$ may not correspond to a connected simplicial lattice. Only those whose indistinguishable open-spin network vertices are connected appropriately (i.e., with appropriate entanglement properties) do \cite{Oriti:2013aqa}. 

From a Fock space perspective, collective (non-perturbative) states, possibly including an arbitrarily large number of GFT quanta, are expected to play a crucial role in extracting continuum physics from GFT. An example of such states is given by coherent states
\begin{equation}\label{eqn:coherentstates}
    \ket{\sigma} = \mathcal{N}_\sigma \exp\left[ \int \diff\vec{g}\diff\boldsymbol{\chi} \, \sigma(\vec{g},\boldsymbol{\chi}) \hat{\varphi}^\dagger(\vec{g},\boldsymbol{\chi}) \right] |0\rangle \, ,
    \quad \vert\mathcal{N}_\sigma\vert^2 = \exp \left[ - \int \diff\vec{g}\diff\boldsymbol{\chi}\, \vert \sigma(\vec{g},\boldsymbol{\chi}) \vert^2 \right],
\end{equation}
which are eigenstates of the annihilation operator $\hat{\varphi}_{\vec{\kappa}(\chi)}$, i.e., $\hat{\varphi}_{\vec{\kappa}(\chi)}\ket{\sigma}=\sigma_{\kappa}(\chi)\ket{\sigma}$. Note, that in virtue of equation \eqref{eqn:closureconstraint}, the wavefunction $\sigma$ is inherently right-diagonal invariant. In addition to that, for concrete computations, $\sigma$ is often assumed to also be left-diagonal invariant,
\begin{equation}
    \sigma(\vec{g},\boldsymbol{\chi})=\sigma(h\cdot_{\text{d}}\vec{g},\boldsymbol{\chi})\,,\qquad\forall h\in\mathrm{SU}(2)\,.
\end{equation}
This invariance is associated with an averaging over the relative embedding of a tetrahedron in $\mathfrak{su}(2)$, and allows the geometric sector of the domain of the condensate wavefunction to be isomorphic to the space of metrics at a point \cite{Gielen:2014ila,Marchetti:2021gcv}.

Coherent states have played a key role in the extraction of cosmological physics from GFTs. Within this context, they are assumed to satisfy not only the properties above, but also a particular form of homogeneity and isotropy, which reduces the wavefunction to the form \cite{Oriti:2016qtz}
\begin{equation}
    \sigma_{\mathbf{\kappa}}(\boldsymbol{\chi})=\sigma_j(\boldsymbol{\chi})(\mathcal{I}^*)^{jjjj,\iota_+}_{m_1m_2m_3m_4}\,,
\end{equation}
where $\iota_+$ is the largest eigenvalue of the volume operator compatible with $j$ and $\boldsymbol{\chi}$ are spatially homogeneous fields. Averages of collective observables (see below) on such homogeneous and isotropic coherent states have indeed been shown to satisfy classical cosmological dynamics with MCMF scalar field matter content \cite{Oriti:2016qtz, Oriti:2016ueo, Gielen:2016dss, Gielen:2019kae, Gielen:2020fgi, Marchetti:2020umh, Jercher:2021bie}. Analogous results have been obtained for slightly inhomogeneous cosmologies, by including quantum entanglement \cite{Jercher:2023kfr, Jercher:2023nxa}. Importantly, these results were obtained by considering a non-interacting GFT, an assumption that will be relaxed in the following. 
\paragraph{Collective operators.}
Within a Fock perspective, the macroscopic dynamics characterizing the emergent phase of the GFT is naturally associated with second-quantized operators. These can be generally written as
\begin{align}\label{eqn:nmbodyobs}
    \hat{O}^{(n,m)} &= \int \left[ \prod_{i=1}^{n} \diff\vec{g}_i \diff\boldsymbol{\chi}_i \, \hat{\varphi}^\dagger(\vec{g}_i,\boldsymbol{\chi}_i) \right] \left[ \prod_{j=1}^m \diff\vec{g}_j^{{}\,\prime} \diff\boldsymbol{\chi}_j' \, \hat{\varphi}(\vec{g}'_j,\boldsymbol{\chi}'_j) \right] O^{(n,m)},
\end{align}
where the matrix elements $O^{(n,m)}$ in principle can depend on all $\vec g_i, \boldsymbol{\chi}_i, \vec{g}_j^{{}\,\prime}, \boldsymbol{\chi}_j'$. 

Some examples of collective operators that will play a key role in the following are the total number and volume operators, as well as the operator corresponding to the (extensive) scalar fields:
\begin{subequations}\label{eqn:exampleobservables}
    \begin{align}\label{eqn:numberoperator}
        \hat{N}&=\int\diff\vec{g}\diff\boldsymbol{\chi} \,\, \hat{\varphi}^\dagger(\vec{g},\boldsymbol{\chi})\hat{\varphi}(\vec{g},\boldsymbol{\chi})\,,\\
        \hat{V}&=\int\diff\vec{g}\diff\vec{g}^{{}\,\prime}\diff\boldsymbol{\chi} \,\, \hat{\varphi}^\dagger(\vec{g},\boldsymbol{\chi})\mathcal{V}(\vec{g},\vec{g}^{{}\,\prime})\hat{\varphi}(\vec{g}^{{}\,\prime},\boldsymbol{\chi})\,, \label{eqn:volumeoperator}\\
        \hat{X}^a&=   \int\diff\vec{g}\diff\boldsymbol{\chi}\,\chi^a\hat{\varphi}^\dagger(\vec{g},\boldsymbol{\chi})\hat{\varphi}(\vec{g},\boldsymbol{\chi})\,,\label{eqn:scalarfieldoperator}
    \end{align}
\end{subequations}
where the kernel $\mathcal{V}(\vec{g},\vec{g}^{{}\,\prime})$ of the volume operator is obtained from the matrix elements of the LQG volume operator on spin-network states \cite{Oriti:2016qtz}.
\paragraph{Relational observables.}
In order to compare the emergent behavior of the GFT model under analysis with the classical one, it is necessary to construct local observables. In classical gravity such a notion of locality can be made compatible with background independence by constructing relational observables, which are localized with respect to a physical (rather than coordinate) reference frame \cite{Goeller:2022rsx}, constituted by dynamical fields. Such a notion of physical locality is natural from a perspective of GFTs, as $\varphi$ is defined on field space. Nonetheless, collective observables such as those in equation \eqref{eqn:exampleobservables} are clearly non-local on field space. This should not be surprising, as equations \eqref{eqn:commutationrelationsg} and \eqref{eqn:commutationrelationsk} do not admit a preferred choice of a physical frame. In this sense, the above Fock structure and the corresponding collective observables are inherently \textit{perspective-neutral} \cite{Marchetti:2024nnk}.

However, it is still possible to define relational observables within the Fock space $\mathcal{F}$. This can be systematically done by introducing a frame POVM and conditioning the above observables with respect to the frame orientation \cite{Marchetti:2024nnk}. Relational observables can also be defined at an effective level by averaging collective observables with respect to some coherent states that are appropriately localized with respect to data of the physical frame. As an example, suppose that the effective localization is performed with respect to a single clock $\chi$. The corresponding localized coherent state would be characterized by a wavefunction of the form
\begin{equation}\label{eqn:peakedcondensate1}
    \sigma_{\epsilon,\pi_0}(\vec{g},\chi,\boldsymbol{\psi})=\eta_\epsilon(\chi-\chi_0,\chimz)\tilde{\sigma}(\vec{g},\chi,\boldsymbol{\psi})\,,
\end{equation}
where $\eta_\epsilon$ is an appropriate peaking function with (small but non-vanishing \cite{Marchetti:2020umh,Marchetti:2020qsq}) width $\epsilon$ around $\chi_0$. A non-vanishing $\pi_0$, associated with a phase $\exp[-\ii\chimz (\chi-\chi_0)]$ was also required in \cite{Marchetti:2020umh} to guarantee a consistent mean-field expansion (see Sec. \ref{sec:variation} below). In the following, we will restrict to the simple case of a Gaussian peaking function,
\begin{equation}
    \eta_{\epsilon} ( \chi - \chiz , \chimz ) 
	= \mathcal{N}_{\varepsilon} \exp{ \left( - \frac{(\chi - \chiz)^{2}}{2 \varepsilon} \right) } \exp{ \left( - \ii \chimz (\chi - \chiz) \right) } 
	\label{eq:etafree}\,.
\end{equation}
Averages of collective observables $\hat{O}$ on the above states $\ket{\sigma_{\epsilon,\psim}}$ become then functionals of the reduced condensate wavefunction $\tilde{\sigma}$ localized at $\chi=\chi_0$:
\begin{equation}\label{eqn:effectiverelational}
    \langle\hat{O}\rangle_{\sigma_{\epsilon,\psim}}=O[\tilde{\sigma}](\chi_0)\,.
\end{equation}
In particular, for the averaged intensive clock $\langle\hat{\chi}\rangle_{\sigma_{\epsilon,\psim}}\equiv \langle\hat{X}\rangle_{\sigma_{\epsilon,\psim}}/\langle\hat{N}\rangle_{\sigma_{\epsilon,\psim}}=\chi_0$ (with $\hat{X}$ being the extensive scalar field operator \eqref{eqn:scalarfieldoperator} associated with $\chi$), showing that $\chi_0$ plays the role of physical time variable. As one would expect, this effective description becomes more and more accurate the smaller $\epsilon$ and the larger $\langle\hat{N}\rangle_{\sigma_{\epsilon,\psim}}$ are. Crucially, in these regimes, the above effective relational observables agree with the POVM-based observables of \cite{Marchetti:2024nnk}. Since in the following we will work in the limit of large average number of quanta and small $\epsilon$, it will be enough to consider effective relational observables of the form \eqref{eqn:effectiverelational}.

\subsection{The GFT action}\label{sec:additionalfield}

As we have mentioned in Sec.\ \ref{sec:introduction}, realistic scalar fields generally possess a non-trivial potential, with significant examples including the Higgs boson and the inflaton. Therefore, it is of considerable physical importance to study their coupling to GFTs, particularly for cosmological applications. On the other hand, self-interacting scalar fields are generally not ideal relational clocks. Even in relatively simple cosmological settings, they fail to provide a global clock, and locally they can lead to non-integrable chaotic dynamics \cite{Hohn:2011us}. For this reason, from now on we will consider two scalar fields: one, $\chi$, minimally coupled, massless and free (MCMF), which we will use as a relational clock, and a second one, $\psi$, with a general non-trivial potential.

As we have seen above, at the level of the underlying GFT, this means that the field $\varphi$ will now depend on both the scalar field data, $\varphi=\varphi(\vec{g},\chi,\psi)$. Analogously, the form of the GFT action
\begin{equation}
	S = K + U + U^{*}
	\label{eq:action}\,,
\end{equation}
and more precisely both the kinetic and interaction kernels $\mathcal{K}$ and $\mathcal{U}$ defining the kinetic $K$ and interaction $U + U^{*}$ terms above, will be affected by the introduction of the interacting field $\psi$. In particular, as it will become clear below, the classical scalar field potential will be encoded in the (T)GFT interaction term $U + \bar{U}$. Below, we will maintain a relatively general analysis by not selecting a specific form for the scalar field potential $V(\psi)$ (and by making only mild assumptions about it), and by restricting only the dependence of the kinetic and interaction kernels on $\psi$.

As reviewed at the beginning of Sec.\ \ref{sec:review}, the GFT action is constructed so that the partition function of the theory generates Feynman amplitudes that can be associated with simplicial matter-gravity path-integrals. As a consequence, if the GFT Lagrangian satisfies particular symmetries, these will be present also at the level of the amplitudes which are associated with the discrete matter-gravity path-integrals. In other words, it is natural to request symmetries of the classical matter-gravity actions to be present also at the level of the GFT action\footnote{However, note that, as we will see explicitly in Sec.\ \ref{sec:emergentscalarfieldcosmology}, there is no guarantee that these symmetries will be present at the level of the emergent, effective dynamics} \eqref{eq:action}. 

At the classical level, the actions for the fields $\chi$ and $\psi$ are given by
\begin{subequations}\label{eqn:classicalscalarfieldactions}
    \begin{align}
        S_\chi&=-\frac{1}{2}\int\diff^4x\sqrt{-g}g^{\mu\nu}\partial_\mu\chi\partial_\nu\chi\\
        S_\psi&=-\frac{1}{2}\int\diff^4x\sqrt{-g}\left(g^{\mu\nu}\partial_\mu\psi\partial_\nu\psi+V(\psi)\right)\,,\label{eqn:classicalinteractingscalaraction}
     \end{align}
\end{subequations}
and are characterized by different symmetries. In particular, $S_\chi$ is invariant under shift $\chi\to\chi+a$ and reflection $\chi\to-\chi$ symmetries, while $S_\psi$, in general, is not. If these classical symmetries are to be preserved by the quantization process, one would expect that the GFT kinetic and interaction kernels are invariant under shift and reflection transformations of the field $\chi$ only. However, this does not exclude the possibility that, for instance, the kinetic kernel alone shows these symmetries for both the fields $\chi$ and $\psi$. In fact, since the kinetic kernel encodes information about the \textit{propagation} of scalar field data between neighboring spacetime cells (usually taken to be $4$-simplices, see below), one would expect that it still shows the symmetries of the (functionally identical) kinetic terms in the classical scalar field actions \eqref{eqn:classicalscalarfieldactions}. Indeed, this is what has been explicitly found in \cite{Li:2017uao}, where the action \eqref{eqn:classicalscalarfieldactions}, discretized over a simplicial complex, has been used to obtain an explicit form for the GFT interaction and kinetic kernels.

More precisely, denoting the difference of the scalar field between vertices $v_{l}$ and $v_{l}'$ connected by the link $l$ by $\chi_{v_{l}} - \chi_{v_{l}'}$, the associated length of the link $L$, the volume $V_{v}$ of the $4$-simplex dual to $v$, as well as the volume $\tilde{V}_{l}$ of the convex $4$-hull of those vertices with the simplex vertices dual to $l$, the discretized action for a self-interacting scalar field $\psi$ becomes \cite{Li:2017uao}
\begin{align}
    S_{\psi}^{T}
    &= \sum_{l} \frac{\tilde{V}_{l}}{2}
    \left(
        \frac{\psi_{v_{l}} - \psi_{v_{l}'}}{L}
    \right)^{2}
    + \sum_{v} V_{v} V(\psi)
    \label{eq:scalarfieldactiondiscretized}\,.
\end{align}
In the same spirit as in Sec.\ \ref{sec:review}, one can compare the discrete scalar field-gravity path-integral with GFT amplitudes to deduce the form of the corresponding GFT action.
The different parts of the coupled group field action are obtained via the $\star$-multiplication of the uncoupled ones with the following factors, read from \eqref{eq:scalarfieldactiondiscretized} \cite{Li:2017uao}:
\begin{align}
    \mathcal{K}_{\psi}
    = \exp{\left(i  \frac{\tilde{V}_{l}}{2} 
    \left(
        \frac{\psi_{v_{l}} - \psi_{v_{l}'}}{L} 
    \right)^{2}\right)}
    \,, \quad
    \mathcal{U}_{\psi} 
    = \exp{\left(\ii V_{v} V(\psi)\right)}
    \label{eq:discretizedactionterms}\,.
\end{align}
The choice of discretizing the scalar fields over the dual nodes means that each simplex must carry a value of the scalar field $\psi$, making the GFT interactions local (in $\psi$). Note that such locality thus follows directly from the properties of the discretized classical scalar field dynamics. Any non-locality could therefore only result from quantum effects (either at the lattice gravity or the non-perturbative GFT level).

\subsubsection{Kinetic term}\label{sec:kinetic}
Based on the argument above, we generalize the GFT kinetic kernel to include both $\chi$ and $\psi$ as follows: 
\begin{align}
	K &= 
	\int \diff \vec{g} \diff \vec{h} \int \diff \chi \diff \chi' \int \diff \psi \diff \psi'\, 
    \varphi^*( \vec{g} , \chi , \psi ) \times
        \nonumber \\
        &\hphantom{{}=\int \diff \vec{g} \diff \vec{h} \int \diff \chi \diff \chi' \int \diff \psi \diff \psi'}
        \times
	\mathcal{K} ( \vec{g} , \vec{h} ; ( \chi - \chi')^{2} ; (\psi - \psi')^{2} ) 
	{\varphi} ( \vec{h} , \chi' , \psi' )
	\label{eq:kineticterm}\,.
\end{align}
As we can see, the kinetic kernel only depends on the squared difference of the two fields $\psi$ and $\psi'$, and thus it is perfectly compatible with the concrete result of \cite{Li:2017uao}. For the concrete computations in Sec.\ \ref{sec:variation}, it will turn out to be useful to Fourier transform with respect to the variable $\psi$. Using Parseval's theorem the kinetic term then becomes
\begin{equation}
	K 
	= \int \diff \vec{g} \diff \vec{h} \int \diff \chi \diff \chi' \int \diff \psim\, 
    \varphi^*( \vec{g} , \chi , \psim ) 
	\mathcal{K} ( \vec{g}, \vec{h} ; ( \chi - \chi')^{2} ; \psim ) 
	{\varphi} ( \vec{h} , \chi' , \psim ) 
	\label{eq:kinetictermfourier}\,,
\end{equation}
where $\pi_\psi$ is the conjugate variable to $\psi$.
Finally, as mentioned at the beginning of the section, in the following we will not need the exact functional form of $\mathcal{K}$, but only its functional dependence. Nonetheless, we will assume (as done in previous works \cite{Oriti:2016qtz,Li:2017uao,Marchetti:2020umh}), that the kinetic kernel admits the following series expansion:
\begin{equation}
	\mathcal{K} ( \vec{g} , \vec{h} ; \chi^{2} ; \psim ) = 
	\sum_{n=0}^{\infty} \frac{\mathcal{K}^{(2n)} ( \vec{g} , \vec{h} ; \psim )}{(2n)!} \chi^{2n}
	\label{eq:kineticexpansion}\,.
\end{equation}
Note that, from a discrete lattice gravity perspective, the GFT kinetic kernel should be considered highly non-local (cfr.\ equation \eqref{eq:discretizedactionterms} and \cite{Li:2017uao}). Thus, the above sum does not in general allow a truncation; as we will see, it is only at the level of the mean-field approximation that such truncations can in fact be meaningful.
\subsubsection{Interaction term}\label{sec:interaction}

GFT interactions are associated with specific cellular structures corresponding to the basic building blocks of the cellular complexes representing discrete spacetime \cite{Krajewski:2011zzu,Freidel:2005qe, Oriti:2011jm}. The properties of these cellular structures are specified by the interaction kernel and by the order $l+1$ of the interaction, i.e., the power of the GFT field (or its complex conjugate) appearing in the interaction term.

A considerable amount of effort has been devoted to the study of two types of interactions. \textit{Simplicial} interactions are characterized by $l+1=5$ powers of the GFT field and by an interaction kernel that pairs field arguments following the combinatorial pattern dictated by the gluing of $5$ tetrahedra to form a $4$-simplex \cite{Krajewski:2011zzu,Freidel:2005qe, Oriti:2011jm}. \textit{Tensor} interactions, on the other hand, are characterized by a symmetry under unitary transformation of the field itself, considered as a rank $d$ (where $d$ is the spacetime dimension) covariant tensor \cite{Pithis:2020kio}. The order of these interactions is arbitrary, but they always include an even number of powers of $\varphi$ and $\bar{\varphi}$ \cite{Pithis:2020kio,deCesare:2016rsf}.

In this work, we will consider a generalization of the above classes of interactions, which, following the terminology of \cite{deCesare:2016rsf}, we call \textit{pseudosimplicial} and \textit{pseudotensorial}. A pseudosimplicial interaction $U_{\text{S}}[\varphi]$ will include an arbitrary power $l+1$ of the GFT field $\varphi$ (respectively, $U^*_{\text{S}}[\varphi^*]$ will include an arbitrary power $l+1$ of the conjugate GFT field $\varphi^*$).  A pseudotensorial interaction $U_{\text{T}}[\varphi,\varphi^*]$, on the other hand, will contain an even (arbitrary) power of the GFT field $\varphi$ and its complex conjugate $\varphi^*$. In both cases, we will leave the form of the kinetic kernels $\mathcal{U}_{\text{S}}$ and $\mathcal{U}_{\text{T}}$ unspecified. However, the kernels $\mathcal{U}_{\text{S}}$ and $\mathcal{U}_{\text{T}}$ will depend on the scalar field potential, $V_\psi(\psi)$, and hence on $\psi$ itself, as shown in \cite{Li:2017uao}. 

Note that considering an arbitrary order $l+1$ of interactions is perfectly natural from the perspective of GFT. Indeed, even if one were to start with a classical GFT action characterized, for example, by simplicial interactions, quantum effects are expected to generate operators of different orders, some of which (potentially different from the simplicial ones) may dominate in a meso- or macroscopic phase of the theory. In other words, an action with generic order $l+1$ could be seen as an effective “intermediate” action, obtained from the UV one by flowing through the theory space. 

\paragraph{Pseudosimplicial interactions.}
For pseudosimplicial interactions, the interaction term can be written as
\begin{equation}
	U_{\text{S}}[\varphi] = \int \diff \chi \diff \psi \int \left( \prod _{a=1}^{l+1} \diff \vec{g}_{a} \right) 
	\mathcal{U}_{\text{S}} (\vec{g}_{1} , \dots , \vec{g}_{l+1} , \psi ) 
	\prod _{a=1}^{l+1} {\varphi} ( \vec{g}_{a} , \chi , \psi )
	\label{eq:interactionterm}\,.
\end{equation}
Note that the kernel $\mathcal{U}_{\text{S}}$ is local with respect to $\psi$: this is because we are assuming the scalar field $\psi$ to be discretized on nodes dual to the cells of the cellular complex used to discretize spacetime. Different choices may lead to a non-local dependence on $\psi$.
Working in Fourier space, the above equation involves a quadruple convolution:
\begin{equation}
	U_{\text{S}} = \int \diff \chi \diff \psim \int \left( \prod _{a=1}^{l+1} \diff \vec{g}_{a} \right)
	\mathcal{U}_{\text{S}} (\vec{g}_{1} , \dots , \vec{g}_{l+1} , \psim )
	\left(
		{\varphi} ( \vec{g}_{1} , \chi ) \ast \dots \ast {\varphi} (\vec{g}_{l+1} , \chi ) 
	\right) (\psim)
	\label{eq:interactiontermfourier}\,.
\end{equation}
It should be noted at this point that the interaction kernel may in general be presented as an operator in the Fourier space of the additional scalar field $\psi$. 
For instance, a potential that is polynomial in $\psi$ appears as a sum of $n$-th derivatives in Fourier space, with $n$ the corresponding powers of $\psi$ in the polynomial;
\begin{equation}
	\psi^{n} \longleftrightarrow \ii^n\frac{\partial^{n}}{\partial \pi_\psi^{n}}
	\label{}\,.
\end{equation}
Polynomial shape of the potential will be considered later -- at this point such an assumption is not necessary.

\paragraph{Pseudotensorial interactions.}
For pseudotensorial interactions, the order $l+1$ is assumed to be even, so that we can write 
\begin{equation}
	U_{\text{T}} = \int \diff \chi \diff \psi \int \left( \prod _{a=1}^{l+1} \diff \vec{g}_{a} \right) 
	\mathcal{U}_{\text{S}} (\vec{g}_{1} , \dots , \vec{g}_{l+1} , \psi ) 
	\prod _{a=1}^{\frac{(l+1)}{2}} {\varphi}^* ( \vec{g}_{a} , \chi , \psi )
	\prod _{a=1}^{\frac{(l+1)}{2}} {\varphi} ( \vec{g}_{a} , \chi , \psi )
	\label{eq:modinteractiontermeven}\,.
\end{equation}
As one can immediately see, these interactions typically do not depend on the phase of the field, and are thus also referred to as \qmarks{modulus interactions}. Also in this case, it turns out to be convenient to work in Fourier space, leading to the pseudotensorial counterpart of equation \eqref{eq:interactiontermfourier}:
\begin{equation}
	U_{\text{T}} = \int \diff \chi \diff \pi_\psi \int \left( \prod _{a=1}^{l+1} \diff \vec{g}_{a} \right) 
	\mathcal{U}_{\text{S}} (\vec{g}_{1} , \dots , \vec{g}_{l+1} , \pi_\psi ) \left(
	\Conv _{a=1}^{\frac{(l+1)}{2}} {\varphi}^* ( \vec{g}_{a} , \chi  )
	\Conv _{a=1}^{\frac{(l+1)}{2}} {\varphi} ( \vec{g}_{a} , \chi )\right)(\pi_\psi)
	\label{eq:modinteractiontermeven}\,,
\end{equation}
where $\Conv _{a=1}^{(l+1)/2}$ represents the convolution product of $(l+1)/2$ factors. Moreover, also in this case we will consider the quantity $\mathcal{U}_{\text{S}}$ as an operator in Fourier space.

\section{Mean-field equations}\label{sec:variation}
Following \cite{Oriti:2016qtz, Marchetti:2020umh,Marchetti:2021gcv}, we will obtain the continuum dynamics via mean-field techniques, representing a (Hilbert space independent) saddle-point approximation of the quantum effective action \cite{Oriti:2021oux}. The mean-field equations can be obtained by imposing the first Schwinger-Dyson equation
\begin{equation}
	0
	= \Braket{ \sigma_{\epsilon \epsilon'} | \frac{\delta S [ \phi , \phi^{\dagger} ]}{\delta \varphi^{\dagger} ( \vec{g} , \chi_0 , \psimz ) } | \sigma_{\epsilon \epsilon'} }
	\label{eq:schwingerdyson}\,,
\end{equation}
where $S$ is the GFT action including either pseudosimplicial or pseudotensorial interactions, and $\ket{\sigma_{\epsilon\epsilon'}}$ is a coherent state as defined in equation \eqref{eqn:coherentstates}, whose condensate wavefunction takes the form
\begin{equation}
	\sigma_{\epsilon \epsilon'}(\vec{g},\chi,\psim) =  \eta_{\epsilon} ( \chi - \chiz , \chimz ) \tilde{\eta}_{\varepsilon'} ( \psim - \psimz  )\tilde{\sigma}(\vec{g},\chi,\psim)
	\label{eq:twicereducedcondensate}\,,
\end{equation}
where $\eta_{\varepsilon} ( \chi - \chiz , \chimz )$ is given by \eqref{eq:etafree}, and $\tilde{\eta}_{\epsilon'}( \psim - \psimz )$ is an appropriate peaking function. Note that, compared to equation \eqref{eqn:peakedcondensate1}, equation \eqref{eq:twicereducedcondensate} includes a peaking function on the conjugate variable  $\psim$ to $\psi$, around\footnote{To avoid potential confusion, we emphasize that $\psimz \neq \pi_{\psi_{0}}$. Moreover, although the exact form of $\tilde{\eta}_{\epsilon'}$ will not be needed in the following, we will assume that $\tilde{\eta}_{\epsilon'}$ has a trivial phase for simplicity.} 
This localization in momentum space, unlike that in $\chi$, is not a requirement related to the definition of effective relational dynamics, but it substantially facilitates the recovery of the correct classical limit in the appropriate regime \cite{Marchetti:2021gcv}, as we will explicitly see in Sec.\ \ref{sec:emergentscalarfieldcosmology}.
\paragraph{Effective mean-field equations.} As can be seen from \eqref{eq:kinetictermfourier}, the averaged variation in \eqref{eq:schwingerdyson} of the kinetic term $K$ is inherently non-local. However, due to the peaking on $\chi = \chiz$ and to the non-trivial phase of $\eta_\epsilon$ (see equation \eqref{eq:etafree}), one can truncate the expansion of the kinetic kernel at second order in the mean-field equations, leading to a corresponding second order differential equation in the clock variable, see \cite{Marchetti:2020umh} for more details. Note that an analogous truncation in $\psim$ is not necessary, nor allowed, due to the trivial phase of $
\tilde{\eta}_\epsilon$.
As a consequence, the resulting variation of the kinetic term of the group field action turns out to be structurally identical to the one performed in \cite{Marchetti:2020umh,Marchetti:2021gcv}, which we refer to for further details.

On the other hand, the variation of the interaction term is inherently different. The $l$-fold convolution appearing in the interaction term \eqref{eq:interactiontermfourier} may seem inconvenient, but it also means that the interaction kernel (which, as emphasized above, may for example be a differential operator) only acts on one of the convolution partners. By pulling the convolution integrals out of the expectation value, one can use the fact that only two of the convolution partners carry the variable $\psim$, together the peaking of the condensate wave function at $\psim = \psimz$, to show that the averaged variation of the interaction term results in just two factors: (a particular combinatorial combination of) the interaction kernel acting on one copy of the conjugate condensate wave function and appropriate powers of the condensate wavefunction and its complex conjugate (see Appendix \ref{a:interaction} for a detailed derivation). 

As discussed in \cite{Oriti:2016qtz}, for EPRL-like interactions, one can show that the  the imposition of condensate wavefunction isotropy results in mean-field equations that are \qmarks{local} in the spin label. Generic interactions will not lead to spin-decoupled mean-field equations. This will be the case, however, if one $j$, say $j_o$ is dominating. Such a single-spin scenario has been shown to naturally emerge in certain models \cite{Gielen:2016uft} and has been assumed in most of the cosmological applications of GFTs. In fact, we will see in Sec.\ \ref{sec:emergentscalarfieldcosmology} that such an assumption is necessary also in this context to match the classical cosmological dynamics in the classical, continuum limit. In the following, we will assume that the equations are local in the spin label, either as a consequence of the properties of the interaction kernels, or because one of the spins is dominating (see Assumption \hyperlink{a2}{A2} below). 
Under these assumptions, and using the above results, the mean-field equations take the following form in spin-representation
\begin{subequations}
\begin{align}
	0 &= \tilde{\sigma}_j'' 
	- 2 \ii \chimzt \tilde{\sigma}_j'- E^{2} \tilde{\sigma}_j
	- \left(\omega_{\text{S},j} \bar{\tilde{\sigma}}_j \right) \bar{\tilde{\sigma}}_j^{l-1}
	\label{eq:groupfieldeomshort}\,,\\
    0 &= \tilde{\sigma}''_j 
	- 2 \ii \chimzt \tilde{\sigma}'_j- E^{2} \tilde{\sigma}_j
	- \left(\omega_{\text{T},j} \bar{\tilde{\sigma}}_j \right) \bar{\tilde{\sigma}}_j^{\frac{l-3}{2}} \tilde{\sigma}_j^{\frac{l+1}{2}}
	\label{eq:groupfieldeomshortpseudotensorial}\,,
\end{align}
\end{subequations}
for pseudosimplicial and pseudotensorial interactions, respectively.
The operators $\omega_{\text{S/T},j}$ are here the spin representation equivalent of $\bar{u}_{\text{S/T}}$, a linear functional of the interaction kernel $\mathcal{U}_{\text{S/T}}$ defined in equation \eqref{eq:smallubar}. 
We also defined for convenience
\begin{equation}
	\chimzt = \frac{\pi_{0}}{\epsilon \pi_{0}^{2} - 1}\,,
	\quad  \quad
	E_{j}^{2} =  \epsilon^{-1} \frac{2}{\epsilon \pi_{0}^{2} - 1} + \frac{B_{j}}{A_{j}}\,,
	\label{eqn:backgroundparameters}
\end{equation}
where $A_j$ and $B_j$ are constants related to the coefficients $\mathcal{K}^{(0)}$ and $\mathcal{K}^{(2)}$ of the kinetic kernel expansion \eqref{eq:kineticexpansion}, see \cite{Marchetti:2020umh} for more details.

As done in \cite{Oriti:2016qtz,Marchetti:2020umh,Marchetti:2021gcv}, we split the equations into real and imaginary parts by decomposing the reduced condensate wavefunction into its modulus and phase, i.e.\ $\tilde{\sigma}_{j} = \rho_{j} \ee^{\ii \theta_{j}}$:
\begin{subequations}\label{eqn:modulusandphaseeqgeneral}
\begin{align}
	0 &= \rho_{j}'' 
	- \left( 
		\left( \theta_{j}' \right)^{2} 
	- 2 \chimzt \theta_{j}' 
	+ E_{j}^{2} 
	\right) \rho_{j}
	- \mathrm{Re} 
	\left[ 
		P^{\text{S/T}}_{j} 
	\right]
	\label{eq:modeom}\,, \\
	0 &= \rho_{j} \theta_{j}''
	+ 2 \rho_{j}' 
	\left( 
		\theta_{j}' - \chimzt
	\right)
	- \mathrm{Im}
	\left[ 
		P^{\text{S/T}}_{j} 
	\right]
	\label{eq:phaseeom}\,,
\end{align}
\end{subequations}
where
\begin{equation}\label{eqn:pj}
\begin{cases}
    P^{\text{S}}_{j} = \omega_{\text{S},j} \left( \rho_{j} \ee^{- \ii \theta_{j}} \right) \rho_{j}^{l-1} \ee^{- l \ii \theta_{j}}\,,\quad &\text{pseudosimplicial}\,,\\
    P^{\text{T}}_{j} = \omega_{\text{T},j} \left( \rho_{j} \ee^{- \ii \theta_{j}} \right) \rho_{j}^{l-1} \ee^{\ii \theta_{j}}\,,\quad &\text{pseudotensorial}\,.
    \end{cases}
\end{equation}
Note that in the non-interacting case, solutions can be easily obtained by making use of appropriate conserved quantities (see equation \eqref{eq:mathcalEbg} below). In the interacting case, however, these quantities are not conserved in general.
Instead, their evolution is closely related to the real and imaginary parts of the interaction term:
\begin{align}
	\tilde{Q}_{j}' 
	&= \rho_{j} \mathrm{Im}
	\left[ 
		P^{\text{S/T}}_j
	\right]
	\label{eq:Qderivative}\,, \\
	\tilde{\mathcal{E}}_{j}' 
	&= 2 \left( 
		\tilde{\mathcal{E}}_{j} - \frac{\tilde{Q}_{j}^{2}}{\rho_{j}^{2}} + \mu_{j}^{2} \rho_{j}^{2} 
	\right)^{\frac{1}{2}}
	\mathrm{Re}
	\left[ 
		P^{\text{S/T}}_j
	\right]
	+ 2 \frac{\tilde{Q}_{j}}{\rho_{j}} 
	\mathrm{Im}
	\left[ 
		P^{\text{S/T}}_j
	\right]
	\label{eq:mathcalEderivative}\,.
\end{align}
Clearly, in order to look for explicit solutions of \eqref{eq:modeom} and \eqref{eq:phaseeom}, one needs to make some assumption on the form of the quantity we denoted by $\omega_j$. From here on, we will assume that $\omega_{\text{S/T},j}$ admits an expansion in $\psi$ of the form
\begin{equation}
	\omega_{\text{S/T},j}(\psi) = \sum_{n} a_{\text{S/T},j}^{(n)} \psi^{n} 
	\xrightarrow{\mathrm{Fourier}}
	\sum_{n} \tilde{a}_{\text{S/T},j}^{(n)} \partial_{\psim}^{n}
	\label{eq:polynomialpotential}\,,
\end{equation}
which in momentum space becomes a differential operator.
While one may be interested also in other non-polynomial functions, this choice is general enough to ideally allow for approximations of such functions, as long as $\omega$ admits a Fourier expansion. 
Note also that any pure gravity interaction that would be present even in the absence of an additional scalar field can be pulled into this polynomial ansatz at the $n=0$ term.
Finally, let us notice that there is in principle no obstruction to a further generalization to higher (even non-specified) numbers of additional fields.

\paragraph{Perturbative setting.}

The equations of motion \eqref{eq:modeom} and \eqref{eq:phaseeom} given the shape of the potential \eqref{eq:polynomialpotential} are partial, nonlinear differential equations. As such, it is a challenging task to find general explicit solutions.
Furthermore, due to the non-invariance of $\tilde{Q}_{j}$ and $\tilde{\mathcal{E}}_{j}$ under relational time translations in the effective Lagrangian it is difficult to reduce the equations to first order differential equations whose solutions may be obtained more easily.
Given these difficulties, we focus on a regime where interactions, although non-negligible, are still small enough to be treated perturbatively. This is the first sensible approximation one can make, while still being able to capture some non-trivial features of the GFT interactions (and thus of the scalar field potential). 
We can then perform a perturbative expansion of the form
\begin{equation}
	\rho_{j} = \bg{\rho} + \varepsilon \fo{\rho} + \mathcal{O}(\varepsilon^{2})
	\quad
	,
	\quad
	\theta_{j} = \bg{\theta} + \varepsilon \fo{\theta} + \mathcal{O}(\varepsilon^{2})
	\label{eq:perturbationmodphase}\,,
\end{equation}
where $\bar{\rho}_j$ and $\bar{\theta}_j$ represent the free (non-interacting) contributions to $\rho_j$ and $\theta_j$, while $\delta \rho_j$ and $\delta\theta_j$ are small correction induced by the correspondingly small GFT interactions. These components are determined by consistently perturbing equations \eqref{eqn:modulusandphaseeqgeneral}. In particular, the background equations are given by
\begin{subequations}\label{eqn:bkgequations}
\begin{align}
	0 
	&= \bg{\rho}'' 
	- \bg{\rho} \left( \bg{\theta}' \right)^{2} 
	+ 2 \chimzt \bg{\rho} \bg{\theta}'
	- E_{j}^{2} \bg{\rho}
	\label{eq:modeombg}\,, \\
	0 
	&= \bg{\rho} \bg{\theta}''
	+ 2 \bg{\rho}' \left( \bg{\theta}' - \chimzt \right)
	\label{eq:phaseeombg}\,,
\end{align}
\end{subequations}
while, at first order, we find
\begin{subequations}\label{eqn:pertequations}
\begin{align}
	0 
	&= \fo{\rho}''
	- 2 \bg{\rho} \bg{\theta}' \fo{\theta}' 
	- \fo{\rho} \left( \bg{\theta}' \right)^{2} 
	+ 2 \chimzt \left( \fo{\rho} \bg{\theta}' + \bg{\rho} \fo{\theta}' \right) 
	- E_{j}^{2} \fo{\rho} 
	- \mathrm{Re} \left[ \bg{P} \right]
	\label{eq:realeomfo}\,, \\
	0 
	&= \bg{\rho} \fo{\theta}'' 
	+ \fo{\rho} \bg{\theta}'' 
	+ 2 \bg{\rho}' \fo{\theta}' 
	+ 2 \fo{\rho}' \left( \bg{\theta}' - \chimzt \right) 
	- \mathrm{Im} \left[ \bg{P} \right]
	\label{eq:imaginaryeomfo}\,.
\end{align}
\end{subequations}
Before proceeding further, let us make two remarks. First, the mean-field perturbative regime considered here still captures non-perturbative effects of the underlying QG dynamics, as the mean-field description includes an infinite sum of Feynman diagrams. Second, one should not expect perturbations to remain small for asymptotically large $\chiz$. In fact, as we will see explicitly below, the interaction terms will eventually grow large enough to dominate over kinetic contributions. 
Thus, even though for any finite clock value one can choose a small enough constant $\varepsilon$ such that the perturbation is valid, it will not be possible to extend our results 
to (positive or negative) arbitrarily large times.

\subsection{Solutions}\label{sec:solutions}
In this section, we will derive explicit solutions to equations \eqref{eqn:bkgequations} (Sec.\ \ref{sec:bkgsol}) and \eqref{eqn:pertequations} (Sec.\ \ref{sec:firstordersolution}), in the cases of pseudosimplicial and pseudotensorial interactions. 
\subsubsection{Background}\label{sec:bkgsol}
Since the background equations of motion are by assumption free, their solutions are identical to those found in \cite{Marchetti:2021gcv}. Here we briefly review how these solutions can be obtained. 
For notational simplicity, from now on we will suppress the explicit dependence of functions on $\chi_0$ and $\psimz$, denoting, for instance $\bg{\rho}\equiv \bg{\rho}(\chi_0,\psimz)$ and $\bg{\theta}\equiv \bg{\theta}(\chi_0,\psimz)$. As a first step, one introduces the quantities
\begin{equation}
	\bg{Q} = \bg{\rho}^{2} \left( \bg{\theta}' - \chimzt \right)
	\,, \qquad
	\bg{\mathcal{E}} = \left( \bg{\rho}' \right)^{2} + \frac{\bg{Q}^{2}}{\bg{\rho}^{2}} - \mu_{j}^{2} \bg{\rho}^{2}
	\,, \label{eq:mathcalEbg}
\end{equation}
which, although conserved, $\bg{Q}'=0=\bg{\mathcal{E}}'$, still depend on $\psimz$. The same applies to 
\begin{equation}
    \mu_{j}^{2} \equiv E_{j}^{2} - \chimzt^{2}\,,
\end{equation}
which crucially determines the asymptotic behavior of the solutions to equation \eqref{eq:modeombg}. These are generically given by \cite{Marchetti:2020umh,Marchetti:2020qsq,Wilson-Ewing:2018mrp}:
\begin{equation}
	\bg{\rho}^{2} 
	= - \frac{\bg{\mathcal{E}}}{2 \mu_{j}^{2}} 
	+ \alpha_{j} \ee^{2 \mu_{j} \chiz} 
	+ \beta_{j} \ee^{- 2 \mu_{j} \chiz}
	\label{eq:modsolbgexact}\,,
\end{equation}
where $\alpha_j$ and $\beta_j$ are integration constants that can be related to $\bar{\mathcal{E}}_j$ and $\bar{Q}_j$. At late (positive) times the second term above ($\sim \ee^{2 \mu_{j} \chiz}$) dominates and thus 
\begin{align}
	\bg{\rho} &\simeq A_{j} \ee^{\mu_{j} \chiz}
	\label{eq:modsolbg}\,, \\
	\bg{\theta} &\simeq \chimzt \chiz - \frac{\bg{Q}}{2 \mu_{j} \bg{\rho}^{2}} + C
	\label{eq:phasesolbg}\,.
\end{align}
Since at late enough times the modulus of the condensate wavefunction is large, we can also write equation \eqref{eq:phasesolbg} as $\bg{\theta}\simeq \chimzt \chiz$. 
Therefore, in the same regime $\bg{\theta}' \simeq \chimzt$ and $\partial_{\psimz} \bg{\theta} \simeq 0$. 
As a consequence, one can easily see that at late times 
\begin{equation}\label{eqn:pider}
    \partial_{\psimz} \left( \bg{\rho} \ee^{\ii \bg{\theta}} \right) \simeq \bg{\rho} \ee^{\ii \bg{\theta}} \left( \partial_{\psimz} \mu_{j} \right) \chiz
\end{equation}
As we will explicitly see in Sec.\ \ref{sec:emergentscalarfieldcosmology}, and as already emphasized in \cite{Marchetti:2020umh}, the classical limit of the cosmological models emerging from the GFT mean-field dynamics is naturally reached in the above late times limit. As a consequence, the form of $\mu_j$ will be strongly constrained by classicality conditions. In particular, it will turn out that in order for the resulting cosmological dynamics to match the general relativistic ones, one would have to assume that $\mu_j$ depends linearly on $\psimz$, at least in the approximate sense that there exists a limit of $\psimz$ in which $\partial_{\psimz}^{2} \mu_{j} \rightarrow 0$. From now on we will assume $\mu_j$ to satisfy this condition.

We will make use of these background solutions to derive solutions to the first order equations \eqref{eqn:pertequations} in the section below.

\subsubsection{First order}\label{sec:firstordersolution}
The first order equations corresponding to the real and imaginary parts of the reduced condensate wave function are in general coupled, and are thus in principle quite challenging to solve. Given equation \eqref{eq:modsolbgexact}, however, we can rearrange equations \eqref{eq:realeomfo} and \eqref{eq:imaginaryeomfo} to obtain third order differential equations for the first order perturbation of modulus and phase:
\begin{subequations}\label{eqn:firstorderexplicit}
\begin{align}
	0 &= \left( 
		\fo{\rho}''' - \mu_{j}^{2} \left( 2 - X_{-}^{2} \right) \fo{\rho}' - \mathrm{Re} \left[ \bg{P}^{\text{S/T}} \right]'
	\right) 
	\nonumber \\
	&\quad+ 3 \mu_{j} X_{-} \left( 
		\fo{\rho}'' - \frac{1}{3} \mu_{j}^{2} \left( 2 X_{-}^{2} + 4 X_{+} - X_{-} X_{+} - 2 \right) \fo{\rho}
		- \mathrm{Re} \left[ \bg{P}^{\text{S/T}} \right]
	\right) 
	\nonumber \\
	&\quad+ \left( \frac{2 \bg{Q}}{\bg{\rho}^{2}} \right)^{2}
	\left(
		\fo{\rho}' - \mu_{j} X_{-} \fo{\rho} - \frac{\bg{\rho}^{2}}{2 \bg{Q}} \mathrm{Im} \left[ \bg{P}^{\text{S/T}} \right]
	\right)
	\label{eq:modeomrearranged}\,, \\
	0 &= \left( 
		\left( \bg{\rho} \fo{\theta} \right)''' 
		- \mu_{j}^{2} \left( 2 - X_{-}^{2} \right) \left( \bg{\rho} \fo{\theta} \right)' 
		- \mathrm{Im}\left[ \bg{P}^{\text{S/T}} \right]'
	\right) 
	\nonumber \\
	&\quad+ 3 \mu_{j} X_{-} \left( 
		\left( \bg{\rho} \fo{\theta} \right)'' 
		- \frac{1}{3} \mu_{j}^{2} 
		\left( 
			6
			- X_{-} X_{+}
			- 2 X_{-}^{2}
		\right) \left( \bg{\rho} \fo{\theta} \right)
		- \mathrm{Im} \left[ \bg{P}^{\text{S/T}} \right]
	\right) 
	\nonumber \\
	&\quad+ \left( \frac{2 \bg{Q}}{\bg{\rho}^{2}} \right)^{2}
	\left(
		\left( \bg{\rho} \fo{\theta} \right)' 
		- \mu_{j} X_{-} \left( \bg{\rho} \fo{\theta} \right) 
		+ \frac{\bg{\rho}^{2}}{2 \bg{Q}} \mathrm{Re} \left[ \bg{P}^{\text{S/T}}\right]
	\right)
	\label{eq:phaseeomrearranged}\,,
\end{align}
\end{subequations}
where $\bg{P}^{\text{S/T}}=P_j^{\text{S/T}}[\bg{\rho}]$. Note that by using equations \eqref{eqn:pj}, \eqref{eq:polynomialpotential} and \eqref{eqn:pider}, we can write
\begin{align}
    \bg{P}^{\text{S/T}}&=\sum_n\mathfrak{a}_{\text{S/T},j}^{(n)}\chiz^n\bg{\rho}^l\times\begin{cases}
        e^{-i(l+1)\bg{\theta}}\,,\qquad &\text{pseudosimplicial}\\
        1\,,\qquad &\text{pseudotensorial}\\
    \end{cases}\\
    \mathfrak{a}_{\text{S/T},j}^{(n)}&\equiv z\tilde{a}_{\text{S/T},j}^{(n)} \left( \partial_{\psimz} \mu_{j} \right) ^n\label{eqn:tildedouble}
\end{align}
Moreover, in equations \eqref{eqn:firstorderexplicit} we defined
\begin{equation}
	X_{\pm} \equiv \frac{\alpha_{j} \ee^{2 \mu_{j} \chiz} \pm \beta_{j} \ee^{- 2 \mu_{j} \chiz}}{\bg{\rho}^{2}}
	\label{eq:definitionX}\,.
\end{equation}
One can see that $X_{\pm} \xrightarrow{\chi \rightarrow \infty} 1$ and thus at sufficiently late times -- which is also the regime where $( 2 \bg{Q} / \bg{\rho}^{2} )^{2}$ is small -- we arrive at the following equations of motion:
\begin{subequations}
\begin{align}
	0 &= 
	\left( 
		\fo{\rho}'' 
		- \mu_{j}^{2} \fo{\rho} 
		- \mathrm{Re} \left[ \bg{P}^{\text{S/T}} \right]
	\right)'
	+ 3 \mu_{j} 
	\left( 
		\fo{\rho}'' 
		- \mu_{j}^{2} \fo{\rho} 
		- \mathrm{Re} \left[ \bg{P}^{\text{S/T}} \right]
	\right)
	\label{eq:feomlate}, \\
	0 &= 
	\left( 
		\left( \bg{\rho} \fo{\theta} \right)'' 
		- \mu_{j}^{2} \left( \bg{\rho} \fo{\theta} \right) 
		- \mathrm{Im} \left[ \bg{P}^{\text{S/T}} \right]
	\right)'
	+ 3 \mu_{j} 
	\left( 
		\left( \bg{\rho} \fo{\theta} \right)'' 
		- \mu_{j}^{2} \left( \bg{\rho} \fo{\theta} \right) 
		- \mathrm{Im} \left[ \bg{P}^{\text{S/T}} \right]
	\right)
	\label{eq:geomlate}.
\end{align}
\end{subequations}
Thus, 
\begin{subequations}\label{eqn:firstorderequationsimpl}
\begin{align}
	\fo{\rho}'' 
	- \mu_{j}^{2} \fo{\rho} 
	- \mathrm{Re} \left[ \bg{P}^{\text{S/T}} \right]
	\simeq C_{j} \ee^{ - 3 \mu_{j} \chiz} 
	&\simeq 0
	\label{eq:modeomlate}\,, \\
	\left( \bg{\rho} \fo{\theta} \right)'' 
	- \mu_{j}^{2} \left( \bg{\rho} \fo{\theta} \right) 
	- \mathrm{Im} \left[ \bg{P}^{\text{S/T}} \right]
	\simeq D_{j} \ee^{ - 3 \mu_{j} \chiz} 
	&\simeq 0
	\label{eq:phaseeomlate}\,,
\end{align}
\end{subequations}
where we have neglected terms $\sim \exp[- 3 \mu_{j} \chiz]$ in the last equality of each line, since at late times they behave as $\sim \bg{\rho}^{-3}$ and thus are largely suppressed.
As a result, in this regime we obtain again second order \textit{decoupled} differential equations, which are clearly simpler to handle. Moreover, it is worth noticing that equations \eqref{eqn:firstorderequationsimpl} are quite similar to one another. This similarity will be particularly useful to find solutions in the concrete cases of pseudosimplicial and pseudotensorial interactions that we consider below.

\paragraph{Pseudosimplicial interactions.}\label{sec:firstordersolutionpseudosimplicial}
In the pseudosimplicial case, using equation \eqref{eq:modsolbg}, the above first-order equations at late times read
\begin{subequations}\label{eq:firstorderimaginarylate}
\begin{align}
	0 &= \foS{\rho}'' 
	- \mu_{j}^{2} \foS{\rho} 
	- \sum_{n} \mathfrak{a}^{(n)}_{\text{S},j}  \alpha_{j}^{l} \chiz^{n} \ee^{l \mu_{j} \chiz} \cos{((l + 1) \bg{\theta})} 
	\label{eq:firstorderlatereal}\,, \\
	0 &= (\bg{\rho} \foS{\theta})'' 
	- \mu_{j}^{2} \bg{\rho} \foS{\theta} 
	+ \sum_{n} {\mathfrak{a}}^{(n)}_{\text{S},j} \alpha_{j}^{l} \chiz^{n} \ee^{l \mu_{j} \chiz} \sin{((l + 1) \bg{\theta})}
	\label{eq:firstorderlateimaginary}\,.
\end{align}
\end{subequations}
Here we have assumed that the constants $\mathfrak{a}^{(n)}_{\text{S}}$ are real for the sake of simplicity. It is important to note, however, that if that was not the case, one could still adopt the same techniques used in these section to find solutions for any given form of the complex coefficients $\mathfrak{a}^{(n)}_{\text{S}}$.

Full solutions to equations \eqref{eq:firstorderimaginarylate} for general $n$ and $l$ are given in terms of incomplete  Gamma functions (see Appendix  \ref{a:gammafunctions}):
\begin{subequations}
\begin{align}
    \foS{\rho} &= 
	\sum_{n} \left( \frac{\mathfrak{a}^{(n)}_{\text{S},j}  \alpha_{j}^{l} \chiz^{n}}{4 \mu_{j} }
	\mathrm{e}^{- \mu_{j} \chiz}
	\left(
		\frac{\Gamma (n+1 , h_{1} \chiz)}{ (h_{1} \chiz)^{n} h_{1}} 
		+ \frac{\Gamma (n+1 , h_{1}^{*} \chiz)}{ (h_{1}^{*} \chiz)^{n} h_{1}^{*}}
	\right) \right.
	\nonumber \\
	&\hphantom{= \sum_{n}}\left.- \frac{\mathfrak{a}^{(n)}_{\text{S},j}  \alpha_{j}^{l} \chiz^{n}}{4 \mu_{j} }
	\mathrm{e}^{\mu_{j} \chiz}
	\left(
		\frac{\Gamma (n+1 , h_{2} \chiz)}{ (h_{2} \chiz)^{n} h_{2}}
		+ \frac{\Gamma (n+1 , h_{2}^{*} \chiz)}{ (h_{2}^{*} \chiz)^{n} h_{2}^{*}}
	\right) \right)
	\label{eq:modsolgeneral}, \\
	\bg{\rho} \foS{\theta} &= 
	\ii \sum_{n} \left( \frac{\mathfrak{a}^{(n)}_{\text{S},j}  \alpha_{j}^{l} \chiz^{n}}{4 \mu_{j} }
	\mathrm{e}^{- \mu_{j} \chiz}
	\left(
		- \frac{\Gamma (n+1 , h_{1} \chiz)}{ (h_{1} \chiz)^{n} h_{1}} 
		+ \frac{\Gamma (n+1 , h_{1}^{*} \chiz)}{ (h_{1}^{*} \chiz)^{n} h_{1}^{*}}
	\right) \right.
	\nonumber \\
	&\hphantom{= \sum_{n}}\left.- \frac{\mathfrak{a}^{(n)}_{\text{S},j}  \alpha_{j}^{l} \chiz^{n}}{4 \mu_{j} } 
	\mathrm{e}^{\mu_{j} \chiz} 
	\left(
		- \frac{\Gamma (n+1 , h_{2} \chiz)}{ (h_{2} \chiz)^{n} h_{2}}
		+ \frac{\Gamma (n+1 , h_{2}^{*} \chiz)}{ (h_{2}^{*} \chiz)^{n} h_{2}^{*}}
	\right) \right)
	\label{eq:phasesolgeneral}.
\end{align}
\end{subequations}
Here, $h_{1} = - (l+1)(\mu_{j} - \ii \tilde{\pi}_{0})$ and $h_{2} = - ((l-1) \mu_{j} - \ii (l+1) \tilde{\pi}_{0})$. The above $\fo{\rho}$ and $\fo{\theta}$ are linear combinations of particular solutions for monomial interaction terms.

As an aside, even though the above equations \eqref{eq:firstorderimaginarylate} and \eqref{eqn:pseudotensorialfirstorderexplicit} can be solved directly, we remark that there is a simpler way to solve them without any further waiving of descriptive power, by looking into separate terms of the non-homogeneous part. 
Instead of solving the first order equations for a general $\bar{n}$-polynomial, one can solve instead the $\bar{n}$ particular equations
\begin{align}
	0 &= \fo{\rho}'' 
	- \mu_{j}^{2} \fo{\rho} 
	- \mathfrak{a}^{(n)}_{\text{S}} A_{j}^{l} \chiz^{n} \ee^{l \mu_{j} \chiz} \cos{((l + 1) \bg{\theta})} 
	\label{eq:firstorderlaterealparticular}\,, \\
	0 &= (\bg{\rho} \fo{\theta})'' 
	- \mu_{j}^{2} \bg{\rho} \fo{\theta} 
	+ \mathfrak{a}^{(n)}_{\text{S}} A_{j}^{l} \chiz^{n} \ee^{l \mu_{j} \chiz} \sin{((l + 1) \bg{\theta})}
    \label{eq:firstorderlateimaginaryparticular}\,.
\end{align}
The full solution can be recovered by adding the particular solutions back together.
It then suffices to focus all following calculations on just one particular set of equations and solutions for some arbitrary $n$.
Since the particular $n$ has been chosen arbitrarily from the available terms of the potential polynomial, we conclude the following: Let $m , n_{1} , n_{2} \in \mathbb{N} \cap \left[ 0 , \bar{n} \right]$ with $m \leq n_{1} \leq n_{2}$.
Note that then the terms corresponding to $k = n_{1} - m$ and $k = n_{2} - m$ in the sums in \eqref{eq:modsolgeneralinteger} and \eqref{eq:phasesolgeneralinteger} differ only in the factors $\mathfrak{a}^{(n_{1})}_{\text{S}}$, $\chiz^{n_{1} - m}$ and $\mathfrak{a}^{(n_{2})}_{\text{S}}$, $\chiz^{n_{2} - m}$ respectively. 
Thus, adding back together the particular solutions reproduces the original potential by simply replacing the factor $\mathfrak{a}^{(n)}_{\text{S}} \chiz^{n}$ by a global sum $\sum_{n} \mathfrak{a}^{(n)}_{\text{S}} \chiz^{n}$.

Note that, in the above equations, fractions of the form $\Gamma(n+1,f \chiz)/(f \chiz)^{n}$ fall off like $\exp[- f \chiz]$ at large $\chiz$ (where $f$ stands for $h_{1}$, $h_{2}$ or their complex conjugates). Moreover, since $n$ is a positive integer, the incomplete Gamma functions can be written as sums over powers of $\chi_{0}$, so that the solutions take the following form:
\begin{subequations}
\begin{align}
	\foS{\rho} 
	&=\! \sum_{n} \frac{\mathfrak{a}^{(n)}_{\text{S}}}{2 \mu_{j}} \bar{\rho}_{j}^{l} n! \mathrm{Re}\left[ \ee^{- i (l+1) \mu_{j} \chi_{0}} \sum_{k=0}^{n} \frac{\chi_{0}^{k}}{k!} \left( h_{1}^{k-n-1} - h_{2}^{k-n-1} \right) \right]
	\label{eq:modsolgeneralinteger}, \\
	\bg{\rho} \foS{\theta} 
	&=\! \sum_{n} \frac{\mathfrak{a}^{(n)}_{\text{S}}}{2 \mu_{j}} \bar{\rho}_{j}^{l} n! \mathrm{Im}\left[ \ee^{- i (l+1) \mu_{j} \chi_{0}} \sum_{k=0}^{n} \frac{\chi_{0}^{k}}{k!} \left( - h_{1}^{k-n-1} + h_{2}^{k-n-1} \right) \right]
	\label{eq:phasesolgeneralinteger}.
\end{align}
\end{subequations}
Each term in the above equations is a combination of sine and cosine functions. For instance, for a given $n$, the contribution to $\fo{\rho}$ corresponding to $k=n$ is given by
\begin{equation}
	\frac{(l+1) \mathfrak{a}^{(n)}_{\text{S},j} \alpha_{j}^{l} \chiz^{n}}{\left| h_{1} \right|^{2} \left| h_{2} \right|^{2}}
	\ee^{l \mu_{j} \chiz}
	\left( 
		2 l \sin{((l+1) \chimzt \chiz)} 
		+ \left( (l-1) \mu_{j}^{2} - (l+1) \chimzt^{2} \right) \cos{((l+1) \chimzt \chi)}
	\right)
	\label{eq:modsolgeneralhighestorder},
\end{equation}
From the above expressions, we can see that neglecting terms suppressed as $ \exp[- 3 \mu_{j} \chiz]$ in \eqref{eq:modeomlate} and \eqref{eq:phaseeomlate} is indeed a consistent approximation, as the above solutions grow as $\exp[l \mu_{j} \chiz] \chiz^{k}$ with $k \geq 0$ for large $\chiz$. 
Finally, note that since
$\tilde{\sigma}_{j} = \bg{\rho} \ee^{\ii \bg{\theta}} + \varepsilon \ee^{\ii \bg{\theta}} \left( \fo{\rho} + \ii \bg{\rho} \fo{\theta} \right)$ at first order, equations \eqref{eq:modsolgeneralinteger} and \eqref{eq:phasesolgeneralinteger} can be combined to obtain, at late times,
\begin{align}
	\tilde{\sigma}_{\text{S},j} 
	&\simeq \bg{\rho} \ee^{\ii \bg{\theta}} 
	\left( 
		1 + \varepsilon \sum_{n} \mathfrak{a}^{(n)}_{\text{S},j}  \ee^{- h_{2} \chiz} 
		\left( 
			\sum_{k = 0}^{n} \frac{n!}{k!} \chiz^{k} \frac{1}{h_{1}^{n-k-1} h_{2}^{n-k-1}} 
			\sum_{m = 0}^{n-k} \left( \frac{h_{2}}{h_{1}} \right)^{m}
		\right)
	\right)
	\label{eq:groupsolgeneralinteger}\,.
\end{align}

\paragraph{Pseudotensorial interactions.}
For pseudotensorial interactions, the first order equations at late times take the form
\begin{align}
	0 &= \foT{\rho}'' - \mu_{j}^{2} \foT{\rho} - \sum_n\mathfrak{a}^{(n)}_{\text{T},j}  \chiz^{n} \alpha_{j}^{l}e^{l\mu_j\chiz}
	\label{eqn:pseudotensorialfirstorderexplicit}\,,\\
    0&=\left( \bg{\rho} \foT{\theta} \right)'' 
	- \mu_{j}^{2} \left( \bg{\rho} \foT{\theta} \right)\,.
\end{align}
The phase equation is easily solved, leading to a constant $\fo{\theta}$. As expected, for pseudotensorial interactions the phase dependence trivializes. The modulus equation can be easily solved by using the same techniques as above, leading to
\begin{equation}
	\foT{\rho} 
	= \sum_{n} \frac{\mathfrak{a}^{(n)}_{\text{T},j} \alpha_{j}^{l} \chiz^{n}}{2 \mu_{j}} 
	\left( 
		\ee^{- \mu_{j} \chiz} \frac{\Gamma(n+1,h_{+}\chiz)}{(h_{+} \chiz)^{n} h_{+}}
		- \ee^{\mu_{j} \chiz} \frac{\Gamma(n+1,h_{-}\chiz)}{(h_{-} \chiz)^{n} h_{-}}
	\right)
	\label{eq:groupmodulusinteractionssol}\,,
\end{equation}
Here, $h_{\pm} = - (l \pm 1) \mu_{j}$, which are simply the real parts of $h_{1}$ and $h_{2}$. Similar to the pseudosimplicial case, since $n$ is a positive integer, we can write
\begin{equation}
	\foT{\rho} 
	= \sum_{n} \frac{\mathfrak{a}^{(n)}_{\text{S}}}{2 \mu_{j}} \bar{\rho}_{j}^{l} n!\sum_{k=0}^{n} \frac{\chi_{0}^{k}}{k!} \left( h_{+}^{k-n-1} - h_{-}^{k-n-1} \right)
	\label{eq:groupmodulusinteractionsolinteger}\,.
\end{equation}
We now have solutions up to first order for both the modulus and phase of the condensate wavefunction, by combining the background solutions \eqref{eq:modsolbg} and \eqref{eq:phasesolbg} with either \eqref{eq:modsolgeneralinteger} and \eqref{eq:phasesolgeneralinteger} for the pseudosimplicial or \eqref{eq:groupmodulusinteractionsolinteger} for the pseudotensorial case.

\section{Observables averages}\label{sec:averages}
In this section, we compute the averages of geometric and matter observables on the coherent states $\ket{\sigma_{\epsilon\epsilon'}}$ with wavefunction given by \eqref{eq:twicereducedcondensate}. Since we are focusing on homogeneous and isotropic configurations, the physics of the system is fully captured by the volume and the scalar field operators defined in equations \eqref{eqn:volumeoperator} and \eqref{eqn:scalarfieldoperator}. As we will see below, the averages of these operators (and their time derivatives) are determined entirely by the solutions obtained in Sec.\ \ref{sec:solutions} for the modulus and phase of the reduced condensate wavefunction.  
\subsection{Volume}

In spin representation, the expectation value of the volume operator on $\ket{\sigma_{\epsilon\epsilon'}}$ is given by
\begin{equation}
	\langle \hat{V}\rangle_{\sigma_{\epsilon\epsilon'}}\equiv  V = \sum_{j} V_{j} \tilde{\sigma}_{j}^{*} \tilde{\sigma}_{j} 
	= \sum_{j} V_{j} \rho_{j}^{2}	\label{eq:volumeexpectation}\,.
\end{equation}
Hence, $V' = 2 \sum_{j} V_{j} \rho_{j} \rho_{j}'$ and $V'' = 2 \sum_{j} V_{j} \left( (\rho_{j}')^{2} + \rho_{j} \rho_{j}'' \right)$. These quantities are crucial to describe the evolution of the Universe, captured by the (relational) Hubble parameter $\mathcal{H}\equiv V'/(3V)$. As $\rho_j$ admits a perturbative decomposition, the same is true for $V=\bar{V}+\delta V$. In the following we will compute $\bar{V}$ and $\delta V$ explicitly. 

\subsubsection{Background contribution}
At the background level, $\bar{V}\equiv\sum_jV_j\bar{\rho}_j^2$ is described in terms of the background modulus solution \eqref{eq:modsolbgexact}. Its derivatives can be easily computed in virtue of the conserved quantities \eqref{eq:mathcalEbg}. In particular, one can straightforwardly obtain \cite{Oriti:2016acw,Marchetti:2020umh,Marchetti:2020qsq}
\begin{subequations}\label{eqn:bkgdynamics}
\begin{align}
	\left(\frac{\bar{V}'}{\bar{V}}\right)^{2} 
	&= 4 
	\left( 
		\frac{
			\sum_{j} V_{j} \bg{\rho} \mathrm{sign}(\bg{\rho}') 
			\left(  
				\bg{\mathcal{E}} 
				- \frac{\bg{Q}^{2}}{\bg{\rho}^{2}} 
				+ \mu_{j}^{2} \bg{\rho}^{2}
			\right)^{\frac{1}{2}}
		}{
			\sum_{j} V_{j} \bg{\rho}^{2}
		}
	\right)^{2}
	\label{eq:bgvoldynamics1}\,, \\
	\frac{\bar{V}''}{\bar{V}}
	&= 4 \frac{
		\sum_{j} V_{j}
		\left( 
			\frac{\bg{\mathcal{E}}}{2}
			+ \mu_{j}^{2} \bg{\rho}^{2}
		\right)
	}{	
		\sum_{j} V_{j} \bg{\rho}^{2}
	}
	\label{eq:bgvoldynamics2}\,,
\end{align}
\end{subequations}
As we will see in Sec.\ \ref{sec:emergentscalarfieldcosmology}, these two quantities are essential for the computation of $\mathcal{H}^2$ and $\mathcal{H}'$, and thus for the comparison of the GFT continuum dynamics with the classical one. 

\subsubsection{First order contribution}
At first order, the perturbed volume is given by $\delta V\equiv 2\sum_jV_j\bar{\rho}_j\delta\rho_j$, and the counterparts of equations \eqref{eqn:bkgdynamics} can be written as
\begin{subequations}\label{eqn:pertvolumedynamics}
\begin{align}\label{eqn:perthubble}
    \delta\left(\left(\frac{V'}{V}\right)^{2}\right) &= 4 \frac{\bar{V}'}{\bar{V}^{3}} \sum_{j} V_{j} \left( \bar{V} \left( \bar{\rho}_{j} \delta\rho_{j}' + \bar{\rho}_{j}' \delta\rho_{j} \right) - \bar{V}' \bar{\rho}_{j} \delta\rho_{j} \right)
    \,,\\
    \delta\left(\frac{V''}{V}\right) &= 2 \frac{1}{\bar{V}^{2}} \sum_{j} V_{j} \left( \bar{V} \left( \bar{\rho}_{j} \delta\rho_{j}'' + \bar{\rho}_{j}'' \delta\rho_{j} + 2 \bar{\rho}_{j}' \delta\rho_{j}' \right) - \bar{V}'' \bar{\rho}_{j} \delta\rho_{j} \right)
    \,.
\end{align}
\end{subequations}
Of course, the above quantities depend on the choice of interaction term in (\ref{sec:interaction}).
For this reason, it is useful to define
\begin{align*}
    G(n,h,\chi) = \frac{\Gamma\left((n+1),h\chi\right)}{\left(h\chi\right)^{n} h}
    \,,\quad h_{\pm} = -(l\pm1)\mu_{j}
    \,,\quad h_{1/2} = h_{\pm} + \ii (l+1)\tilde{\pi}_{0}
    \,.
\end{align*}
and distinguish between pseudosimplicial and pseudotensorial interactions.
\paragraph{Pseudosimplicial interactions.}\label{sec:volumedynamicspseudosimplicial}
From \eqref{eq:modsolgeneral} it is straightforward to calculate the necessary derivatives of the modulus solution:
\begin{subequations}\label{eqn:rhodersolsimpl}
\begin{align}
    \foS{\rho} &= \sum_{n} \frac{\mathfrak{a}^{(n)}_{\text{S},j} \chi_{0}^{n}}{4 \mu_{j}} \bar{\rho}_{j}^{l} \left( \ee^{h_{+} \chi_{0}} \left( G(n,h_{1}) + G(n,h_{1}^{\ast}) \right) - \ee^{h_{-} \chi_{0}} \left( G(n,h_{2}) + G(n,h_{2}^{\ast}) \right) \right)
    \,,\\
    \foS{\rho}' &= - \mu_{j} \foS{\rho} - \sum_{n} \frac{\mathfrak{a}^{(n)}_{\text{S},j} \chi_{0}^{n} A^{l-1}}{2} \bar{\rho}_{j} \left( G(n,h_{2}) + G(n,h_{2}^{\ast}) \right)
    \,,
\end{align}
\end{subequations}
where the second derivative can be obtained from the dynamical equation \eqref{eq:firstorderlatereal}.

\paragraph{Pseudotensorial interactions.}
Analogously, from equation \eqref{eq:groupmodulusinteractionssol}, we obtain
\begin{subequations}\label{eqn:rhodersoltens}
\begin{align}
    \foT{\rho} &= \sum_{n} \frac{\mathfrak{a}^{(n)}_{\text{T},j} \chi_{0}^{n}}{2 \mu_{j}} \bar{\rho}_{j}^{l} \left( \ee^{h_{+}} G(n,h_{+}) - \ee^{h_{-}} G(n,h_{-}) \right)
    \,,\\
    \foT{\rho}' &= - \mu_{j} \delta\rho_{j} - \sum_{n} \mathfrak{a}^{(n)}_{\text{T},j} \chi_{0}^{n} A^{l-1} \bar{\rho}_{j} G(n,h_{-})
    \,.
\end{align}
\end{subequations}
Again, the second derivative can be obtained from the dynamical equation \eqref{eqn:pseudotensorialfirstorderexplicit}.

\subsection{Matter}\label{sec:matterdynamics}
In Sec.\ \ref{sec:emergentscalarfieldcosmology} we will study the emergent dynamics for the scalar field $\psi$, and we will compare it to the classical ones. To this purpose, in this section we will compute the expectation value on the coherent state $\ket{\sigma_{\epsilon\epsilon'}}$ of the \textit{extensive} scalar field observable $\hat{\Psi}$, given by equation \eqref{eqn:scalarfieldoperator}, $\Psi\equiv \langle\hat{\Psi}\rangle_{\sigma_{\epsilon\epsilon'}}$. As for the volume, $\Psi$ will admit a decomposition into a background $\bar{\Psi}$ and perturbative $\delta\Psi$ components, which we write explicitly below.

\subsubsection{Background contribution}\label{sec:matterdynamicsbg}
As for the volume, the background computations are identical to those of \cite{Marchetti:2021gcv}, which we briefly review here. 
Given the late-time solutions for the background modulus and phase, one can calculate the expectation value of the scalar field operator as
\begin{align}
	\bar{\Psi} 
	&= \bar{N} \sum_{j} \partial_{\psimz} \bg{\theta}
	\nonumber \\
	&= \sum_{j} \left(\frac{\bg{Q}}{\mu_{j}} \partial_{\psimz} \mu_{j} \chiz
	- \frac{1}{2} \partial_{\psimz} \left( \frac{\bg{Q}}{\mu_{j}} \right)
	+ \bar{N} \partial_{\psimz} c_j \right)
	\nonumber \\
    &\equiv \bar{\psi}+\Delta\bar{\Psi}\,,
	\label{eq:bgscalarfield}\,,
\end{align}
where $\bar{N}$ is the background component of the average number of quanta $N\equiv \langle\hat{N}\rangle_{\sigma_{\epsilon\epsilon'}}$, and where
\begin{equation}
    \bar{\psi}\equiv \sum_{j} \left(\frac{\bg{Q}}{\mu_{j}} \partial_{\psimz} \mu_{j} \chiz
	- \frac{1}{2} \partial_{\psimz} \left( \frac{\bg{Q}}{\mu_{j}} \right)\right)\,,\qquad \Delta\bar{\Psi}\equiv \sum_j\bar{N} \partial_{\psimz} c_j \,.
\end{equation}
Note that $\Delta\bar{\Psi}$ is proportional to $\bar{N}$, and thus a purely extensive contribution to $\bar{\Psi}$. This is not the case for $\bar{\psi}$, which is instead just a linear function of $\chi_0$. As we are interested only in the intensive component of the average scalar field, we will choose from here on the constants $c_j$ so that $\Delta\bar{\Psi}=0$.

\subsubsection{First order contribution}\label{sec:matterdynamicsfo}
The perturbative contribution to the expectation value of the scalar field operator is 
\begin{equation}
	\delta \Psi 
	= \sum_{j} \left( \delta N (\partial_{\psimz} \bg{\theta}) 
	+ \bar{N} (\partial_{\psimz} \fo{\theta}) \right)\equiv \delta\psi+\Delta\Psi
	\label{eq:fomatterexpectation}\,,
\end{equation}
where $\delta N$ is the first order contribution to $N$, and where we have defined
\begin{subequations}
\begin{align}
	\delta\psi&\equiv\left( \delta N (\partial_{\psimz} \bg{\theta}) \right)''
	= 
	2 \left( 
		\frac{\fo{\rho}''}{\bg{\rho}} 
		- 2 \mu_{j} \frac{\fo{\rho}'}{\bg{\rho}}
		+ \mu_{j}^{2} \frac{\fo{\rho}}{\bg{\rho}}
	\right) \bar{\psi}
	+ 4 \left( 
		\frac{\fos{\rho}'}{\bgs{\rho}}
		- \mu_{j_{o}} \frac{\fos{\rho}}{\bgs{\rho}}
	\right) \bar{\psi}'
	\label{eq:fomatterexpectationfirst}\,,\\
    \Delta\Psi&\equiv \left( \bar{N} \partial_{\psimz} \fo{\theta} \right)''
	= 
	\bar{N} \left( 
		4 \mu_{j}^{2} \left( \partial_{\psimz} \fo{\theta} \right)
		+ 4 \mu_{j} \left( \partial_{\psimz} \fo{\theta} \right)'
		+ \left( \partial_{\psimz} \fo{\theta} \right)'' 
	\right)
    \,.
    \label{eq:fomatterexpectationsecond}
\end{align}
\end{subequations}
As before, we note that $\Delta\Psi$ is in fact an extensive contribution to $\Psi$. Contrarily to the background case, however, $\Delta\Psi$ cannot be put to zero by simply choosing appropriate integration constants. Note that this term is only relevant in the case of pseudosimplicial interactions, as the phase then has a nontrivial first order contribution. Like with the modulus in the previous section, it is straightforward to calculate the necessary derivatives:
\begin{subequations}
\begin{align}
    \foS{\theta} &= \ii \sum_{n} \frac{\mathfrak{a}^{(n)}_{\text{S},j} \chi_{0}^{n}}{4 \mu_{j}} \bar{\rho}_{j}^{l-1} \left( \ee^{h_{+} \chi_{0}} \left( - G(n,h_{1}) + G(n,h_{1}^{\ast}) \right) - \ee^{h_{-} \chi_{0}} \left( - G(n,h_{2}) + G(n,h_{2}^{\ast}) \right) \right)
    \,,\\
    \foS{\theta}' &= - \ii \sum_{n} \frac{\mathfrak{a}^{(n)}_{\text{S},j} \chi_{0}^{n} A_{j}^{l+1}}{2} \bar{\rho}^{-2} \left( - G(n,h_{2}) + G(n,h_{2}^{\ast}) \right)
    \,,
\end{align}
\end{subequations}
and the second derivative via the dynamics \eqref{eq:firstorderlateimaginary}.

We now have collected all necessary expressions for the volume and matter dynamics both at the background (equations \eqref{eqn:bkgdynamics} and \eqref{eq:bgscalarfield}) and first order (equations \eqref{eqn:pertvolumedynamics} and \eqref{eq:fomatterexpectation}) level, as well as for all derivatives of the modulus and phase that appear in the former. With this, one can continue with the physical analysis.

\section{Emergent scalar field cosmology}\label{sec:emergentscalarfieldcosmology}
In this section, we will explore under which conditions and in which regimes the interacting GFT dynamics matches the classical cosmological ones. To this end, we will compare the relational evolution of the average volume and matter field computed in Section \ref{sec:averages} with the corresponding cosmological general relativistic dynamics. The latter will be evaluated in the harmonic gauge corresponding to the choice of an MCMF scalar field as a clock \cite{Oriti:2016qtz,Gielen:2018fqv}.

As one would expect, the underlying GFT dynamics, even after a mean-field approximation, generally describes a pre-geometric, non-spatiotemporal physics. Therefore, to recover a continuum geometric limit, additional approximations are necessary. 
\begin{description}
    \item[A1]\hypertarget{a1} As discussed in \cite{Marchetti:2020qsq,Gielen:2019kae}, a continuum and classical limit of the mean-field GFT dynamics is naturally associated with a regime of large $\rho_j$ and, consequently, a large average number of quanta $N=\sum_j\rho_j$. As seen from the solution \eqref{eq:modsolbgexact}, this regime can be reached (irrespective of the initial conditions) at sufficiently large relational times. In this limit, quantum fluctuations of observables remain well-controlled. Furthermore, the effective relational framework developed above is consistent \cite{Marchetti:2020qsq}. In the following, we will therefore restrict our analysis to this regime. However, it is important to note that, since we are assuming GFT interactions to be perturbatively small, the analysis cannot be extended to arbitrarily large times, as interactions will eventually become dominant (see also Sec.\ \ref{sec:variation}).
\end{description}
Even though relational observables become effectively classical in this limit and can be used to describe continuum geometries, their effective dynamics will not in general coincide with the corresponding classical ones. This matching can only be obtained in specific regimes:
\begin{description}
    \item[A2]\hypertarget{a2} As shown in \cite{Oriti:2016qtz,Marchetti:2020umh}, even in the classical limit described above, the effective GFT dynamics can match the cosmological, general relativistic dynamics only if one of the modes of the condensate wavefunction, say $j=j_o$, dominates. Since interactions are included perturbatively here, this condition remains necessary to achieve a cosmological matching at the background level, where the field $\psi$ is effectively MCMF. Consequently, in the following, we will restrict our analysis to this \qmarks{single-spin} scenario.
\end{description}
In addition to the above assumptions—which, as already mentioned, can be justified from the perspective of a non-interacting GFT—we will adopt the following additional assumptions regarding the interaction terms. These will significantly simplify the comparison with classical cosmological physics.
\begin{description}
    \item[A3]\hypertarget{a3} We assume that the GFT interaction kernel $\omega_{\text{S/T},j}$ depends \textit{linearly} on the scalar field potential $V_{\psi}$, i.e., 
    \begin{equation}\label{eqn:linearapproxw}
        \omega_{\text{S/T},j}(\psi)\equiv\omega_{\text{S/T},j}(V_\psi)=\omega_{\text{S/T},j}(0)+\frac{\partial \omega_{\text{S/T},j}}{\partial V_\psi}\biggr\vert_{V_\psi=0} V_\psi\,.
    \end{equation}
    Note that, from a simplicial gravity perspective, this is generally not the case (see \cite{Li:2017uao} and equation \eqref{eq:discretizedactionterms}), as the potential takes the form $\sim \exp[V_vV_\psi]$, where $V_v$ is the volume of a $4$-simplex $v$. However, within this framework, it seems natural to assume that the $4$-simplices making up the spacetime discretization are of Planckian size. Consequently, in the classical, low-energy regime—where one might expect the quantum GFT dynamics to approximate the classical, general relativistic ones—the scalar field should not attain trans-Planckian values. In this regime, $V_vV_\psi$ remains small, and the linear approximation \eqref{eqn:linearapproxw} is indeed justified. 
    \item[A4]\hypertarget{a4} Finally, we assume the scalar field $\psi$ is evaluated on-shell \cite{Marchetti:2021gcv} at the level of interactions. Since $\psi$-interactions are encoded exclusively in GFT interactions—and thus appear only as first-order corrections in our perturbative framework—any difference between $\psi$ and $\chi_0$ at this level is negligible. Indeed, at zeroth order, $\bar{\psi}'=\mathfrak{p}$, where $\mathfrak{p}\equiv \Pi_\psi/\Pi_\chi$ by construction. To simplify further the following analysis, we will assume that the shift between $\bar{\psi}$ and $\chiz$ is negligible (although technically non-zero, for the reasons explained in \cite{Marchetti:2021gcv}) i.e., $\bar{\psi}\simeq \mathfrak{p} \chi_0$. 
\end{description}
In this work, we will retain the assumptions outlined above. However, we will discuss potential directions for relaxing them in Sec.\ \ref{sec:conclusion}.
\subsection{Background contribution}
As we will see below, it will be natural to compare the emergent GFT dynamics with the evolution of a Universe whose matter content consists of a MCMF scalar field $\chi$, a self-interacting scalar field $\psi$ with potential $V_\psi$, and a (small) dark energy component $\Lambda$. At the classical level, the dynamics of such a system are given by the following equations:
\begin{subequations}\label{eqn:classicaleqs}
\begin{align}
    \left( 
		\frac{V'}{3 V}
	\right)^{2}
	&= \frac{8 \pi G}{3} 
	\left( 
		1
		+ \frac{\Pi_{\psi}^{2}}{\Pi_{\chi}^{2}}
		+ \frac{V^{2}}{\Pi_{\chi}^{2}} (\Lambda +V_{\psi})
	\right)
	\label{eq:classicalfriedmann1}\,,\\
     0 &= \psi'' + \frac{V^{2}}{\Pi_{\chi}^{2}} \frac{\diff V_{\psi}}{\diff \psi}\,,
\end{align}
\end{subequations}
Note that the above equations are written in the (harmonic) gauge corresponding to the choice of the field $\chi$ as a clock, so that $'\equiv \diff/\diff\chi$, $\mathcal{H}\equiv ({V}'/(3V))$ and $V\equiv V_{\text{fid}}a^3$, with $a$ the scale factor and $V_{\text{fid}}$ the fiducial volume. Furthermore, $\Pi_{\chi}$ and $\Pi_{\psi}$ are constants (which can be identified with the free scalar field momenta). Note that under the assumption of small $\Lambda$ and $V_\psi$, one can perform the following perturbative decomposition of the above equations. At the background level, we find 
\begin{equation}
    \left( \frac{\bar{V}'}{3\bar{V}} \right)^{2}=\frac{8 \pi G}{3} 
	\left( 
		1
		+ \frac{\Pi_{\psi}^{2}}{\Pi_{\chi}^{2}}\right)\,,\qquad \bar{\psi}''=0\,,
\end{equation}
In this section, we will see under which conditions the background (free) GFT evolution of the average volume and matter field can be recast in the above form.
\paragraph{Volume dynamics.}
Within the late-time (\hyperlink{a1}{A1}) and single-spin (\hyperlink{a2}{A2}) approximations above, equations \eqref{eq:bgvoldynamics1} and \eqref{eq:bgvoldynamics2} become
\begin{equation}
	\left(\frac{\bar{V}'}{\bar{V}}\right)^{2} 
	= 4 \mu_{j_{o}}^{2}
	= \frac{\bar{V}''}{\bar{V}}
	\label{eq:bgvoldynamicslimit}\,.
\end{equation}
The above equations can be straightforwardly compared with equations \eqref{eq:classicalfriedmann1} in the limit in which $V_\psi=0=\Lambda$, i.e.
\begin{equation}
    \left(\frac{\bar{V}'}{\bar{V}}
	\right)^{2}
	= 24\pi G
	\left( 
		1
		+ \frac{\Pi_{\psi}^{2}}{\Pi_{\chi}^{2}}
	\right)=\frac{\bar{V}''}{\bar{V}}\,.
\end{equation}
As already discussed in \cite{Marchetti:2021gcv}, it is natural in this framework to associate $\psimz$ with the unperturbed momentum $\Pi_\psi$ of $\psi$, and $\chimzt$ as the momentum $\Pi_\chi$ of $\chi$, in which case the above equations match only if $\mu_{j_o}(\psimz)$ satisfies
\begin{equation}\label{eqn:mumatching}
    \mu^2_{j_o}(\psimz)=6\pi G(1+\psimz^2/\chimzt^2)\,,
\end{equation}
at least in an appropriate regime. As an example, note that if the field $\psi$ dominates the background evolution, so that $\psimz^2/\chimzt^2\gg 1$, a linear dependence $\mu_{j_o}\simeq \psimz/\chimzt$ would still allow to match the classical dynamics \cite{Marchetti:2021gcv}.
\paragraph{Matter dynamics.}
A similar matching can be performed for the average matter field dynamics. However, since the details of this matching will not be important in the following, we will not repeat the analysis here, but we will simply remind that matching the classical and quantum effective dynamics at the background level results on imposing appropriate conditions on the parameter $\bgs{Q}$ \cite{Marchetti:2021gcv}. Further details can be found in \cite{Marchetti:2021gcv}.

To summarize, following the procedure outlined in \cite{Marchetti:2021gcv} and reviewed here, the GFT effective dynamics can, in an appropriate classical limit, match the classical evolution of a universe containing minimally coupled, massless, free scalar fields. In the next section, we will study under which conditions the perturbed GFT dynamics can yield a volume evolution that incorporates small contributions from both a scalar field potential and a dark energy component.
\subsection{Including interactions}
At first order in the small potential $V_\psi$ and cosmological constant $\Lambda$, the classical equations \eqref{eqn:classicaleqs} take the form
\begin{align}
    \delta \left(\left( \frac{V'}{3V} \right)^{2}\right) &=\frac{8\pi G}{3}\frac{\bar{V}^2}{\Pi^2_\chi}(\Lambda+V_\psi)\label{eqn:perturbedfriedmannclassical}\\
    \delta\psi''&=-\frac{\bar{V}^{2}}{\Pi_{\chi}^{2}} \frac{\diff V_{\psi}}{\diff \psi}\,\label{eqn:perturbedscalarfield}
\end{align}
Here we will compare the above equations to the perturbed volume and matter GFT effective dynamics, distinguishing, as usual, between the pseudo-tensorial and the pseudo-simplicial case.
\subsubsection{Pseudotensorial interactions}\label{sec:matchingtens}
Under Assumption \hyperlink{a1}{A1}, one can approximate the general solution \eqref{eq:groupmodulusinteractionssol} as
\begin{equation}\label{eqn:latetimestensorial}
    \foT{\rho}\simeq\frac{\bar{\rho}_j^{l}}{\mu_j^2(l^2-1)}\sum_n\mathfrak{a}_{\text{T},j}^{(n)}\chi^n_0\equiv \frac{\bar{\rho}_j^{l}}{\mu_j^2(l^2-1)}(\tilde{\Lambda}+\mathcal{V})\,,
\end{equation}
where $\tilde{\Lambda}=\mathfrak{a}_{T,j}^{(0)}$. As we will always work within the late-times limit of Assumption \hyperlink{a1}{A1}, $\simeq$ symbols will be replaced by $=$ symbols throughout.
\paragraph{Volume dynamics.} 
Using equation \eqref{eqn:latetimestensorial} in equation \eqref{eqn:perthubble}, together with Assumptions \hyperlink{a2}{A2}, \hyperlink{a3}{A3}, and \hyperlink{a4}{A4}, we can see that the perturbed Hubble parameter takes the form
\begin{equation}\label{eqn:perturbedhubbletens}
    \delta\left(\left(\frac{V'}{3V}\right)^2\right)=\frac{8}{9}\frac{1}{l+1}\frac{\bar{V}^{(l-1)/2}}{V_{j_o}^{(l-1)/2}}(\tilde{\Lambda}+\mathcal{V})
\end{equation}
where we have also assumed that $\vert\mathcal{V}'\vert\ll \vert \mu_{j_{o}}\mathcal{V}\vert$ for simplicity. Comparing this with the classical equation \eqref{eqn:perturbedfriedmannclassical},
we see that the two equations can be matched provided that $l=5$ and that
\begin{subequations}
\begin{align}
    \tilde{\Lambda}&=\Lambda 18 \pi G \frac{V_{j_o}^2}{\Pi_\chi^2}\,,\\
    \mathcal{V}&=V_\psi 18 \pi G \frac{V_{j_o}^2}{\Pi_\chi^2}\,.\label{eqn:curlyv}
\end{align}
\end{subequations}
In turn, the quantities $\tilde{\Lambda}$ and $\mathcal{V}$ can be related to the Fourier transform of the GFT kinetic kernel using equations \eqref{eqn:tildedouble} and \eqref{eqn:mumatching}. As we can see, under the assumptions outlined at the beginning of this section, order $6$ pseudo-tensorial interactions produce a continuum volume dynamics consistent with that of a universe containing two minimally coupled scalar fields—one free and the other weakly interacting—and a small dark energy component. Interestingly, similar types of GFT interactions have been (phenomenologically) explored in other works by starting with an effective interaction kernel, and constraining its form by comparing the cosmological results with the actual evolution of the universe. Order $6$ pseudo-tensorial interactions turned out to be necessary, within that phenomenological framework, to recover an accelerated expansion both at early times  (inflation)\footnote{Note, however, that order $6$ pseudo-tensorial interactions alone are not enough to provide a graceful exit mechanism for the inflationary phase, which is the reason why in \cite{deCesare:2016rsf} interactions of different orders were phenomenologically combined. } \cite{deCesare:2016rsf}, or at late times (dark energy) \cite{Oriti:2025lwx}.
\paragraph{Matter dynamics.}
We can follow the same procedure for the scalar field dynamics. Using equation \eqref{eqn:latetimestensorial} in equation \eqref{eq:fomatterexpectationfirst}, together with Assumptions \hyperlink{a2}{A2}, \hyperlink{a3}{A3}, and \hyperlink{a4}{A4}
\begin{equation}\label{eqn:perturbedmattertens}
    \delta\psi''=2\frac{l-1}{l+1}\bar{\rho}_{j_o}^{l-1}(\tilde{\Lambda}+\mathcal{V})\bar{\psi}\overset{l=5}{=}\frac{4}{3}\frac{\bar{V}^2}{V_{j_o}^2}(\tilde{\Lambda}+\mathcal{V})\bar{\psi}\,.
\end{equation}
This should be compared with the classical equation \eqref{eqn:perturbedscalarfield}. We note that in general, the quantum equation is characterized by an \textit{emergent mass} term, given by
\begin{equation}
    m^2_\psi=-\frac{4}{3}\frac{\tilde{\Lambda}}{V_{j_o}^2}\,.
\end{equation}
which is positive only if $\tilde{\Lambda}<0$ and vice versa. This shows how a negative cosmological constant can produce particle masses, effectively breaking any shift symmetry present at the classical level, and thus realizing a \textit{quantum-gravitational counterpart of the Higgs mechanism}. This phenomenon was already suggested in anti-deSitter space (see e.g.\ \cite{Porrati:2001db}). Besides this novel emergent mass term, the scalar field equations of motion can be matched only if 
\begin{equation}\label{eqn:matchingscalarpotential}
    -\frac{\bar{V}^{2}}{\Pi_{\chi}^{2}} \frac{\diff V_{\psi}}{\diff \bar{\psi}}=\frac{4}{3}\frac{\bar{V}^2}{V_{j_o}^2}\mathcal{V}\bar{\psi}=24\pi G\frac{\bar{V}^2}{\Pi_\chi^2}V_\psi\bar{\psi}\,,
\end{equation}
where in the last equation we used \eqref{eqn:curlyv}. This can only be satisfied if
\begin{equation}
    V_\psi=V_0\exp[-\bar{\psi}^2(12\pi G)]\,,
\end{equation}
suggesting that the matching, in general, is only possible for a specific form of the potential. Note that $V_\psi\le V_0$, so if $V_0$ is small enough to guarantee the above perturbative decomposition, this requirement will be satisfied for any value of $\bar{\psi}$. Moreover, note that the resulting potential is \textit{technically natural}, as in the decoupling limit $G\to 0$ the scalar field becomes subject to a constant potential, and the resulting action is invariant under shift symmetry $\psi\to\psi+c$.

Finally, let us observe that the condition $\vert\mathcal{V}'\vert\ll \vert \mu_j\mathcal{V}\vert$ yields, for the above solution, $\vert \bar{\psi}24\pi G\mathfrak{p}/\mu_{j_o}\vert\ll 1$, i.e., $\vert\bar{\psi}\vert\ll 2\sqrt{(1+\mathfrak{p}^2)/(6\pi G\mathfrak{p}^2)}$, which is certainly satisfied if the field $\bar{\psi}$ is not trans-Planckian. Nonetheless, one can repeat the above analysis relaxing the assumption that $\vert\mathcal{V}'\vert\ll \vert\mu_{j_o}\mathcal{V}\vert$, at the expenses of a more complicated counterpart of equation \eqref{eqn:matchingscalarpotential}. Indeed, the volume matching condition will now relate 0th and 1st order derivatives of $\mathcal{V}$ with $V_\psi$, while the matter matching condition will relate derivatives up to the second order of $\mathcal{V}$ to $\diff V_\psi/\diff\psi$. These equations can be combined to obtain a single equation for $\mathcal{V}$, which in turn will fix the form of $V_\psi$. 
\paragraph{Running couplings.}
Finally, let us observe that the above matching can be satisfied for \textit{any} form of the potential, provided that the gravitational coupling is appropriately renormalized. Indeed, in general one would expect that the quantum gravity interactions may produce a non-trivial running of the emergent couplings. As interactions are introduced perturbatively, this phenomenon is expected to affect only couplings present at the level of the background dynamical equations, i.e., $G$, $\Pi_\chi$ and $\Pi_\psi$. Within a cosmological context, running can be naturally defined with respect to the clock itself. For this reason, since $\chi$ and $\psi$ are dynamically indistinguishable at the background level, one may expect both $\Pi_\chi$ and $\Pi_\psi$ not to run. Even if that was the case, however, the background dynamical indistinguishability of $\psi$ and $\chi$ would suggest that the ratio between their momenta (i.e., $\mathfrak{p}$) is in fact unaffected by the renormalization flow.

Therefore, here we will assume that only the gravitational coupling is running with respect to the physical clock $G(\chiz)\equiv \bar{G}+\delta G(\chiz)$. As a consequence, equation \eqref{eqn:curlyv} becomes
\begin{equation}
    \mathcal{V}=V_\psi 18\pi \bar{G}\frac{\bar{V}_{j_o}^2}{\bar{\Pi}_\chi^2}+\frac{8\pi\delta G}{3}(1+\mathfrak{p}^2)\,,
\end{equation}
so that one can obtain a consistent matching between the quantum equations and the renormalized classical ones, by requiring
\begin{equation}\label{eqn:deltagtens}
    \delta G=-\left(\frac{\diff V_\psi}{\diff\bar{\psi}}+12\pi\bar{G}V_\psi\bar{\psi}\right)\left(\frac{8\pi\bar{\Pi}_\chi^2}{3\bar{V}^2_{j_o}}\bar{\psi}(1+\mathfrak{p}^2)\right)^{-1}.
\end{equation}
This suggests that, from the perspective of perturbatively interacting quantum gravity, renormalization group (RG) flow of the classical parameters can be obtained by requiring a consistent matching between the emergent quantum-gravitational dynamics, and the RG-improved classical ones.
The running of emergent gravitational couplings comes with (or rather from) a running of the GFT couplings. While imposed somewhat post-hoc here, it should be possible to study them moving beyond the mean field approximation, by performing again the derivation of effective cosmological dynamics from the quantum effective action (possibly truncated at one loop).

\subsubsection{Pseudosimplicial interactions}
The analysis for the pseudo-simplicial case follows closely the one outlined in Sec.\ \ref{sec:matchingtens} above. Once again, using Assumption \hyperlink{a1}{A1}, we approximate the general solution \eqref{eq:modsolgeneral} as 
\begin{equation}\label{eqn:latetimessimplicial}
    \foS{\rho}=\bar{\rho}_{j}^l\sum_n\mathfrak{a}^{(n)}_{\text{S},j}\chiz^n\mathrm{Re}\left(\frac{e^{-i(l+1)\chimzt\chiz}}{h_1h_2}\right)\equiv \bar{\rho}_{j}^l(\tilde{\Lambda}+\mathcal{V})F_j(\chimzt\chiz)\,,
\end{equation}
where $\tilde{\Lambda}$ and $\mathcal{V}$ are defined as in \eqref{eqn:latetimestensorial}, and where $F_j$ can be written as
\begin{equation}
    F_{j}=\frac{\cos((l+1)\chimzt\chiz+\vartheta)}{\vert h_1 h_2 \vert}\,,\qquad h_1h_2=\vert h_1h_2\vert e^{\ii \vartheta}\,,
\end{equation}
where one should remember that $h_{1}$, $h_{2}$, and $\vartheta$ all implicitly depend on depend on $j$.
\paragraph{Volume dynamics.}
Once again, using \eqref{eqn:latetimessimplicial} in equation \eqref{eqn:perthubble}, together with Assumptions \hyperlink{a2}{A2}, \hyperlink{a3}{A3}, and \hyperlink{a4}{A4}, we can see write the perturbed Hubble parameter takes the form
\begin{equation}\label{eqn:hubblepertsimplicialapp}
    \delta\left(\left(\frac{V'}{3V}\right)^2\right)=\frac{8}{9}\mu_{j_{o}}\frac{\bar{V}^{2}}{V_{j_o}^{2}}\left(F_{j_o}'+4\mu_{j_o}F_{j_o}\right)(\tilde{\Lambda}+\mathcal{V})\,,
\end{equation}
where, as before, we assumed that $\mathcal{V}$ varies slowly compared to $F_{j_o}$ and $\bar{\rho}_{S,j_{o}}^l$. Comparing equations \eqref{eqn:perturbedhubbletens} (with $l=5$) and \eqref{eqn:hubblepertsimplicialapp}, we see that they differ by the trigonometric factor $(F_{j_o}'+4\mu_{j_o}F_{j_o})\mu_{j_o}$. Thus, equation \eqref{eqn:curlyv} changes accordingly:
\begin{equation}\label{eqn:curlyvsimp}
    \mu_{j_{o}}\left(F_{j_o}'+4\mu_jF_{j_o}\right)\mathcal{V}=3\pi G\frac{V_{j_o}^2}{\Pi_\chi^2}V_\psi\,.
\end{equation}
The trigonometric function $F_{j_o}$ remains nearly constant for field values such that $\vert \bar{\psi} / f_\mathfrak{p} \vert \ll 1$, where $f_\mathfrak{p} \equiv \vert \mathfrak{p} / ((l+1)\chimzt) \vert$. In this limit, equation \eqref{eqn:hubblepertsimplicialapp} reduces to \eqref{eqn:perturbedhubbletens} (with $l=5$), so that the same arguments provided in the previous section hold. 

The scale $f_\mathfrak{p}$ can be determined as follows. Since $\chimzt^2 = E_{j_o}^2 - \mu^2_{j_o}$, in the limit $\epsilon \chimz^2 \gg 1$ (a requirement for the effective relational framework to be well defined \cite{Marchetti:2020qsq}), one can express $\chimzt^2$ as $\chimzt^2 = \mu^2_{j_o} + B_{j_o} / A_{j_o}$. This means that if $\vert B_{j_o} / A_{j_o} \vert$ is subdominant, $f_\mathfrak{p}$ is of the order $\mu_{j_o}^{-1}$. Using equation \eqref{eqn:mumatching}, we have $\mu_{j_o}^2 = 6\pi G(1 + \Pi^2_\psi / \Pi^2_\chi) = 6\pi G(1 + \mathfrak{p}^2)$, and thus, in this regime, we can write
    \begin{equation}
        f_\mathfrak{p}= M_{\text{Pl}}\frac{\vert \mathfrak{p}\vert}{(l+1)\sqrt{6\pi (1+\mathfrak{p}^2)}}< \frac{M_{\text{Pl}}}{\sqrt{6\pi}(l+1)}\,,
    \end{equation}
where the upper limit is quickly reached as soon as $\mathfrak{p}\gtrsim 1$. This represents the more realistic regime where the matter field of interest ($\psi$) dominates over the clock field ($\chi$). In this limit, the corrections introduced by $F_{j_o}$ are pushed to the Planck scale, highlighting their inherently quantum gravitational nature.

Importantly, note that if the potential $V_\psi(\psi)$ is constant, $V_\psi=V^{(0)}$, the classical theory possesses a shift symmetry $\psi \to \psi + c$. Quantum gravitational effects break this symmetry down to the discrete subgroup $\psi \to \psi + 2\pi n / f_\mathfrak{p}$, with $n$ an integer. The scale $f_\mathfrak{p}$ thus acts as the symmetry-breaking scale. This is analogous to how non-perturbative effects in axions break the classical shift symmetry \cite{GrillidiCortona:2015jxo}, although, in this case, the symmetry-breaking effects are quantum gravitational in origin and arise from the coupling of the scalar field to gravity rather than to a gauge field.
 
Finally, note that when $l=5$ and $V_\psi=0$ (or even in the more extreme case where the field $\psi$ is entirely absent), equation \eqref{eqn:hubblepertsimplicialapp} yields a contribution to the Hubble parameter that is compatible with that of a dark energy component, corrected by the trigonometric function $F_{j_o}$ (which by construction depends on $\chi_0$ if $\psi$ is absent). Notably, these contributions have profoundly rich phenomenological implications, spanning scenarios from natural inflation \cite{Adams_1993,PhysRevLett.65.3233} to evolving dark energy \cite{Copeland:2006wr,Caldwell:2009ix,Kamionkowski:2014zda}. Importantly, within our framework, these contributions are generated \textit{naturally} by quantum gravitational interactions.
\paragraph{Matter dynamics.} As in the case of tensorial interactions, whether $\mathcal{V}$ gives rise to a quantity that can be associated with a scalar field potential depends on how that very same quantity enters the equation of motion for the scalar field. To determine this we can proceed as in the previous section, with two important differences. These findings will be discussed further in \cite{Ladstatter:2025abc}.

First, the perturbed scalar field \eqref{eq:fomatterexpectation} contains in this case terms $\sim \bar{N}\delta\theta_{S,j_o}$. From equations \eqref{eq:phasesolgeneralinteger}, we can see that these terms scale as $\bar{\rho}_{j_{o}}^{l+1}$. As recognized in \cite{Marchetti:2021gcv}, and emphasized in Sec.\ \ref{sec:matterdynamicsfo}, they are associated with extensive behavior, and can be seen as a consequence of the inherently extensive definition of the scalar field operator \eqref{eqn:scalarfieldoperator}. In particular, for $l=5$ they do not allow a natural matching with the classical dynamics. We thus remove their contribution by hand, by \textit{restricting} the scalar field perturbation to its intrinsic part 
\begin{equation}
    \delta\psi=\sum_j\delta N(\partial_{\psimz}\bar{\theta}_j)\,.
\end{equation}
A second difference from the tensorial case is given by the fact that, when $\vert\bar{\psi}/f_\mathfrak{p}\vert \gtrsim 1$, the quantity $\mathcal{V}$ will not be associated directly with $V_\psi$, as the trigonometric function $F{_{j_o}}$ now enters the matching condition \eqref{eqn:curlyvsimp}. Indeed, the intensive contribution to $\delta\psi''$ is given by
\begin{align}\label{eqn:pertmattersimp}
    \delta\psi''&=2\frac{\bar{V}^2}{V_{j_o}^2}\left(-36F_{j_o}\beta_{j_o}+8\mu_{j_o}F_{j_o}'\right)(\tilde{\Lambda}+\mathcal{V})\bar{\psi}\,,
\end{align}
where $r_{j_o}=B_{j_o}/A_{j_o}$. The matching condition then becomes
\begin{equation}\label{eqn:matchingscalarpotentialsimp}
    \frac{\diff V_\psi}{\diff\bar{\psi}}=6\pi G V_\psi\mathcal{F}_{j_o}\bar{\psi}\,,\qquad\mathcal{F}_{j_o}\equiv \frac{-36F_{j_o}r_{j_o}+8\mu_{j_o}F_{j_o}'}{\mu_{j_{o}}(F_{j_o}'+4\mu_jF_{j_o})}
\end{equation}
We can proceed as in Sec.\ \ref{sec:matchingtens} and integrate the above equation to obtain a scalar field potential that allows a consistent matching with the classical dynamics. For instance, this can be easily done when $\vert r_{j_o}\vert\ll 1$ (hence $\mu_{j_o}\simeq \chimzt$), in which case
\begin{equation}
    \mathcal{F}_{j_o}=12\left[1-\frac{2}{3}\cot((l+1)\chimzt\chiz+\vartheta))\right]^{-1}\,,
\end{equation}
although the final expression for $V_\psi$ is not particularly illuminating. 

As in the previous case, a consistent matching is possible for an arbitrary form of the potential as long as the gravitational coupling is renormalized appropriately. Comparing equation \eqref{eqn:matchingscalarpotential} with \eqref{eqn:matchingscalarpotentialsimp}, it is straightforward to obtain the counterpart of equation \eqref{eqn:deltagtens}:
\begin{equation}\label{eqn:deltagsimplicial}
    \delta G=-\left(\frac{\diff V_\psi}{\diff\bar{\psi}}+12\pi\bar{G}\mathcal{F}_{j_o}V_\psi\bar{\psi}\right)\left(\frac{8\pi\bar{\Pi}_\chi^2}{3\bar{V}^2_{j_o}}\bar{\psi}(1+\mathfrak{p}^2)\right)^{-1}.
\end{equation}

To summarize, we now have first order expressions for the volume (equations \eqref{eqn:perturbedhubbletens} and \eqref{eqn:hubblepertsimplicialapp}) and matter (equations \eqref{eqn:perturbedmattertens} and \eqref{eqn:pertmattersimp}) dynamics in both pseudotensorial and pseudosimplicial scenarios, together with appropriate matching and running conditions. 

\section{Conclusions and discussion}\label{sec:conclusion}
In this work, we studied the cosmological dynamics emerging from GFT models coupled to a minimally coupled, massless, free scalar field—serving as a relational clock—and a self-interacting scalar field. The effects of scalar field interactions were encoded in a local dependence of the GFT interaction kernel on the scalar field itself. We focused on two broad classes of GFT interactions that generalize simplicial and tensorial interactions: those depending on an arbitrary power of the GFT field $\varphi$ alone, and those depending on an even arbitrary power of both $\varphi$ and its complex conjugate $\varphi^*$. These interactions were denoted as pseudosimplicial and pseudotensorial, respectively.

For these two classes of interactions, we derived mean-field equations by averaging the quantum equations of motion on appropriately chosen coherent peaked states, following the procedure outlined in \cite{Marchetti:2020umh,Marchetti:2021gcv}. Assuming that the interaction kernel is local in representation space and that the interactions are subleading, we then perturbatively analyzed the equations of motion and found explicit solutions at first order in interactions. We subsequently focused on averages (on the aforementioned coherent peaked states) of geometric and matter observables, represented respectively by the volume and scalar field operators, and derived their effective relational dynamics using the explicit solutions to the mean-field equations.

We then compared the effective dynamical equations for the relational volume and matter scalar field with the corresponding classical equations in the harmonic gauge, which corresponds to choosing a minimally coupled, massless, and free scalar field clock. In doing so, we assumed: (i) a late-time regime (large number of quanta); (ii) a single-spin scenario; (iii) a linear dependence of the interaction kernel on the scalar field potential; and (iv) that the interacting scalar field can be evaluated on-shell at first order in interactions. In the geometric sector, a matching with the classical Friedmann dynamics is possible for a specific order of interactions in both the pseudotensorial and pseudosimplicial cases. However, in the former, a cosmological constant term emerges from quantum gravity interactions, while in the latter, this emergent component takes the form of dynamical dark energy.

In a Higgs-mechanism-like fashion, the same quantum gravity interactions also introduce an effective mass term in the scalar field dynamics. This term is constant (and of opposite sign to the cosmological constant) in the pseudotensorial case, and time-dependent in the pseudosimplicial case. In both cases, we have shown that a geometry-compatible classical matching in the matter sector is only possible for specific forms of the effective scalar field potential. In particular, we highlighted that, in the pseudotensorial case, such potentials are technically natural, while in the pseudosimplicial case, quantum gravity effects intrinsically alter the symmetry group of the original classical potential. For instance, we noted that a shift symmetry of the classical potential is naturally broken to a discrete subgroup, which resembles non-perturbative effects in axion physics \cite{GrillidiCortona:2015jxo}. Finally, we highlighted that these quantum gravity-compatibility conditions on the effective potentials can be relaxed by allowing the gravitational coupling to run, showing that such running can be determined uniquely once a choice of the classical potential is made.

The emergence of a dark energy component in pseudotensorial GFT models is already a well-known result \cite{deCesare:2016rsf,Oriti:2025lwx}. However, the associated mass generation demonstrated in this work is a genuinely novel phenomenon, as it suggests a quantum gravity–induced Higgs mechanism for self-interacting scalar fields. It would be important to explore whether a similar mechanism exists for fermions if they are coupled to the underlying GFT. We leave this, and the study of the intriguing physical consequences of this phenomenon to future work. Additionally, we have shown that even within a single-mode scenario, the emergent dark energy component for pseudosimplicial interactions is dynamical. This quantum gravity–induced acceleration could provide both a mechanism for triggering inflation without an inflaton and a way to physically motivate dynamical dark energy models that may help alleviate cosmological tensions. These possibilities will be investigated in future work \cite{Ladstatter:2025abc}.

Even if an inflationary scenario within these models does not require an inflaton, it is important to note that if $\psi$ is to be interpreted as an inflaton, the quantum gravity–compatible potentials for both pseudotensorial and pseudosimplicial interactions do not appear to be favored by current observations \cite{Planck:2018jri}. However, we remark that these results were obtained perturbatively and thus may not be reliable if the scalar field potential becomes too large, which is typically the case in inflationary models. Nonetheless, if these results hold beyond the perturbative regime, we emphasize that a single-field inflationary scenario compatible with current observations would still be possible through an appropriate running of the gravitational coupling. Notably, equations \eqref{eqn:deltagtens} and \eqref{eqn:deltagsimplicial} suggest that $G$ becomes constant again after reheating.

Let us remind the reader that such a time-dependent gravitational coupling was introduced to ensure the validity of the classical equations of motion regardless of the form of the scalar field potential. This time dependence of $G$ is inherently linked to the presence of quantum gravity interactions, making it natural to interpret it as a consequence of its renormalization group (RG) flow. If this interpretation is confirmed, the matching procedure outlined in this work could provide an effective, bottom-up approach to deriving RG flow equations in physically relevant scenarios. This would be particularly valuable given that the top-down alternative—performing a full RG analysis of the quantum gravity system and examining the \textit{relational} running of macroscopic couplings—is currently out of reach. Nonetheless, it remains crucial to test this intuition by comparing bottom-up and top-down approaches in simple toy models.

Finally, in future works it would be important to relax some of the simplifying assumptions made here. In particular, for most phenomenological studies—especially those concerning inflation and dynamical dark energy—it is necessary to go beyond the perturbative regime considered here \cite{Ladstatter:2025abc}. Additionally, it would be interesting to explore the physical consequences of relaxing the single-mode assumption \hyperlink{a2}{A2} by including a second mode, following the approach of \cite{Oriti:2025lwx}.
\acknowledgments{
    We thank Daniele Oriti for useful comments on a first draft of the manuscript, as well as Renata Ferrero and Kristina Giesel for helpful discussions. L.M.~acknowledges support from the Kavli Insitute for the Physics and Mathematics of the Universe, from the Okinawa Institute of Science and Technology Graduate University, and from the John Templeton Foundation, through ID\# 62312 grant as part of the \href{https://www.templeton.org/grant/the-quantum-information-structure-of-spacetime-qiss-second-phase}{\textit{`The Quantum Information Structure of Spacetime'} Project (QISS)}.}
\appendix
\section{Variation of the interaction term}\label{a:interaction}
This is a more detailed treatment of the interaction term \eqref{eq:interactionterm} as it enters the Schwinger-Dyson equations \eqref{eq:schwingerdyson}. While this describes the pseudosimplicial case in particular, the exact same procedure applies to the pseudotensorial one. Thus, we will omit the S/T subscript in the equaitons below, for simplicity. First, Fourier transform leads to a convolution in the integrand, i.\ e.\
\begin{align}
    U &= \int \diff\chi \diff\psi \int\left(\prod_{a=1}^{l+1}\diff \vec{g}^{a}\right) \mathcal{U}(\vec{g}^{1},\dots,\vec{g}^{l+1},\psi) \prod_{a=1}^{l+1} \varphi(\vec{g}^{a},\chi,\psi)
    \nonumber\\
    &= \int \diff\chi \diff\pi_{\psi} \delta(\pi_\psi) \int\left(\prod_{a=1}^{l+1}\diff \vec{g}^{a}\right) \int\left(\prod_{b=1}^{l+1}\diff \pi_{b}\right) \left( \mathcal{U} \ast \varphi(\vec{g}^{1}, \chi) \ast \dots \ast\varphi(\vec{g}^{l+1},\chi) \right)(\pi_\psi)
    \nonumber\\
    &=\int \diff\chi \diff\pi_{\psi} \delta(\pi_\psi) \int\left(\prod_{a=1}^{l+1}\diff \vec{g}^{a}\right) \int\left(\prod_{b=1}^{l+1}\diff \pi_{b}\right)
    \times\nonumber\\
    &\hphantom{=} \times \mathcal{U}(\vec{g}^{1},\dots,\vec{g}^{l+1},\pi_{\psi}-\pi_{1}) \varphi(\vec{g}^{1},\chi,\pi_{1}-\pi_{2}) \dots \varphi(\vec{g}^{l+1},\chi,\pi_{l+1})
    \nonumber\\
    &= \int \diff\chi \int\left(\prod_{a=1}^{l+1}\diff \vec{g}^{a}\right) \int\left(\prod_{b=1}^{l+1}\diff \pi_{b}\right) 
    \times\nonumber\\
    &\hphantom{=} \times \mathcal{U}(\vec{g}^{1},\dots,\vec{g}^{l+1},-\pi_{1}) \varphi(\vec{g}^{1},\chi,\pi_{1}-\pi_{2}) \dots \varphi(\vec{g}^{l+1},\chi,\pi_{l+1})
    \,.
\end{align}
Next, find the functional derivative
\begin{align}
    \frac{\delta \bar{U}}{\delta \varphi^{\dagger}(\vec{g},\chi,\psi)} &= \int \diff\chi \int\left(\prod_{a=1}^{l+1}\diff \vec{g}^{a}\right) \int\left(\prod_{b=1}^{l+1}\diff \pi_{b}\right) \left( \mathrm{[I]} + \mathrm{[II]} + \dots + \mathrm{[l+1]} \right)
    \,,\\
    \mathrm{[I]}&=
        \delta(\pi-(\pi_{1}-\pi_{2})) \delta(\vec{g}-\vec{g}^{1}) \left( \tilde{\bar{\mathcal{U}}} \varphi^{\dagger} \right) (\vec{g}^{2},\chi,\pi_{2}-\pi_3) \varphi^{\dagger} (\vec{g}^{3},\chi,\pi_{3}-\pi_4) \dots 
        \nonumber\\
        &\hphantom{=} \dots \varphi^{\dagger} (\vec{g}^{l+1},\chi,\pi_{l+1})
    \,,\nonumber\\
    \mathrm{[II]}&=
        \left( \bar{\mathcal{U}} \varphi^{\dagger} \right)(g_{1},\chi,\pi_{1}-\pi_{2}) \delta(\pi-(\pi_{2}-\pi_{3})) \delta(\vec{g}-\vec{g}^{2}) \varphi^{\dagger} (\vec{g}^{3},\chi,\pi_{3}-\pi_4) \dots 
        \nonumber\\
        &\hphantom{=} \dots \varphi^{\dagger} (\vec{g}^{l+1},\chi,\pi_{l+1})
    \,,\nonumber\\
    \dots
    \,,\nonumber\\
    \mathrm{[l+1]}&=
        \left( \bar{\mathcal{U}} \varphi^{\dagger} \right)(g_{1},\chi,\pi_{1}-\pi_{2}) \varphi^{\dagger} (g_{2},\chi,\pi_{2}-\pi_{3}) \dots 
        \nonumber\\
        &\hphantom{=} \dots \varphi^{\dagger} (\vec{g}^{l},\chi,\pi_{l}-\pi_{l+1}) \delta(\pi-\pi_{l+1}) \delta(\vec{g}-\vec{g}^{l+1})
    \nonumber
    \,,
\end{align}
where $\bar{\mathcal{U}} = \bar{\mathcal{U}}(\vec{g}^{1},\dots,\vec{g}^{l+1},-\pi_{1})$ and $\tilde{\bar{\mathcal{U}}} = \bar{\mathcal{U}}(\vec{g}^{1},\dots,\vec{g}^{l+1},-\pi_{2})$.
We need to keep in mind that $\mathcal{U}$ (or rather $\bar{\mathcal{U}}$) may act as a differential operator in momentum space. In particular, for a polynomial shape we have
\begin{equation}\nonumber
    \sum_{n} c_{n} \psi^{n} \longleftrightarrow \sum_{n} \tilde{c}_{n} \partial_{\pi_{\psi}}^{n} 
\end{equation}
Clearly this operator only acts on one of the $\varphi$, and it is the reason the first term had to be integrated by parts, leading to one instance of $\tilde{\bar{\mathcal{U}}}$.
Evaluating the integrals affected by delta distributions, relabeling the $\pi_{b}$ and $\vec{g}^{a}$, and defining
\begin{equation}\label{eq:smallubar}
    \bar{u}(\vec{g}_1,\dots,\vec{g}^l,\pi_1) = \bar{\mathcal{U}}(\vec{g},\vec{g}^{1},\dots,\vec{g}^{l},-\pi_{1}) + \bar{\mathcal{U}}(\vec{g}^{1},\vec{g},\vec{g}^{2},\dots,\vec{g}^{l},-\pi_{1}) + \dots + \bar{\mathcal{U}}(\vec{g}^{1},\dots,\vec{g}^{l},\vec{g},-\pi_{1})
    \,,
\end{equation}
simplifies the functional derivative to
\begin{align}
    \frac{\delta \bar{U}}{\delta \varphi^{\dagger}(\vec{g},\chi,\psi)} 
    &= \int \diff \chi \int\left(\prod_{a=1}^{l}\diff \vec{g}^{a}\right) \int\left(\prod_{b=1}^{l}\diff \pi_{b}\right) \left( \bar{u} \varphi^{\dagger} \right) (\vec{g}^{1},\chi,-\pi-\pi_{1})
    \times \nonumber\\
    &\hphantom{=} \times \varphi^{\dagger} (\vec{g}^{2},\chi,\pi_{1}-\pi_{2}) \dots \varphi^{\dagger}(\vec{g}^{l},\chi,\pi_{l})
    \,.
\end{align}
Finally, the expectation value of the above variation on coherent peaked states
\begin{subequations}
\begin{align}
    \sigma_{\epsilon \epsilon'}(\vec{g},\chi,\pi) &= \tilde{\sigma}(\vec{g},\chi,\pi)
    \eta_{\epsilon}(\chi-\chi_{0},\pi_{\chi 0}) \eta_{\epsilon'}(\pi-\pi_{\psi 0},\psi_{0})
    \,,\\
    \eta_{\epsilon}(\chi-\chi_{0},\pi_{0}) &= \mathcal{N}_{\epsilon} \exp{\left( - \frac{(\chi - \chi_{0})^{2}}{2 \epsilon} \right)} \exp{\left( - \ii \pi_{0} (\chi - \chi_{0}) \right)}
    \,,\\
    \eta_{\epsilon'}(\pi-\pi_{\psi 0},\psi_{0}) &= \mathcal{N}_{\epsilon'} \exp{ \left( - \frac{(\pi - \pi_{\psi 0})^{2}}{2 \epsilon'} \right) } \exp{ \left( -\ii \psi_{0} (\pi - \pi_{\psi 0}) \right) }
    \,,
\end{align}
\end{subequations}
is approximated as
\begin{align}
    \simeq& \int\left(\prod_{a=1}^{l}\diff \vec{g}^{a}\right) \left( \bar{u} \left( \tilde{\sigma}(\vec{g}^{1},\chi,-\pi-(l-1)\pi_{\psi 0}) \eta_{\epsilon'}(-\pi-l\pi_{\psi 0},\psi_{0}) \right) \right) \prod_{b=1}^{l-1} \tilde{\sigma}(\vec{g}^{b},\chi_{0},\pi_{\psi 0})
    \nonumber\\
    \simeq& \int\left(\prod_{a=1}^{l}\diff \vec{g}^{a}\right) \eta_{\epsilon'}(-\pi-l\pi_{\psi 0},\psi_{0}) \left( \bar{u} \tilde{\sigma} \right)(\vec{g}^{1},\chi,-\pi-(l-1)\pi_{\psi 0}) \prod_{b=1}^{l-1} \tilde{\sigma}(\vec{g}^{b},\chi_{0},\pi_{\psi 0})
    \,.
\end{align}
Note that peaking in the momentum of $\psi$ happened at $\pi_{\psi 0}$, while the functional derivative above was taken at some arbitrary value $\pi$. The remaining peaking function effectively enforces  $\pi=-l\pi_{\psi 0}$, so that the above expression leads to the last term in \eqref{eq:groupfieldeomshort} in spin representation.

\section{Incomplete Gamma Functions}\label{a:gammafunctions}
The Gamma function is defined as the analytic continuation of
\begin{equation}
	\Gamma(z)
	= \int_{0}^{\infty} \diff x x^{z-1} \ee^{-x}
	\label{def:gamma}\,.
\end{equation}
The upper Gamma function $\Gamma(s,n)$ and lower Gamma function $\gamma(s,n)$ are defined similarly, but with different, incomplete, integral boundaries:
\begin{align}
	\Gamma(s,x) &= \int_{x}^{\infty} t^{s-1} \ee^{-t} \diff t
	\label{def:uppergamma}\,, \\
	\gamma(s,x) &= \int_{0}^{x} t^{s-1} \ee^{-t} \diff t
	\label{def:lowergamma}\,.
\end{align}
Intuitively, they represent the split of the Gamma function at a specified value $s$.
In the context of this work, the variable $x$ is a (linear) function, which we will now adapt here as well for clarity as $f(x)$.
For positive integer values $s=n$ the upper Gamma function simplifies to
\begin{equation}
	\Gamma(n,f(x)) = n! \ee^{-f(x)} \sum_{k=0}^{n-1} \frac{f^{k}(x)}{k!}
	\label{eq:uppergammainteger}\,.
\end{equation}
One can easily see how for large $n$ or small $f(x)$ it approaches the factorial. In fact, one can show that even in the case of noninteger values $s$ for the asymptotic behaviour:
\begin{equation}
	\frac{\Gamma(s,f(x))}{f^{s-1}(x) \ee^{-f(x)}} \xrightarrow{f(x) \to \infty} 1
	\label{eq:uppergammalimit}\,.
\end{equation}
Luckily, the derivative of the upper Gamma function is rather simple:
\begin{equation}
	\frac{\diff }{\diff x} \Gamma \left( s , f(x) \right) 
	= - \ee^{ - f(x) } \left(f(x)\right)^{s-1} f'(x)
	\label{eq:uppergammaderivative}\,.
\end{equation}

\bibliographystyle{JHEP}
\bibliography{references-new.bib}

\end{document}